\documentclass[pre,twocolumn,amsmath,amssymb]{revtex4}

\usepackage{graphicx}
\usepackage{dcolumn}
\usepackage{bm}

\begin{document}

\newcommand{\vspfigA}{\vspace{0.23cm}}  
\newcommand{\vspfigB}{\vspace{0.25cm}}  
\newcommand{\widthfigA}{0.35\textwidth} 
\newcommand{\widthfigB}{0.41\textwidth}  
\newcommand{\widthfigC}{0.44\textwidth}  
\newcommand{\widthfigD}{0.85\textwidth}  

\baselineskip 0.455cm



\title{Boundary effects in the stepwise structure of the Lyapunov spectra   
for quasi-one-dimensional systems}

\author{Tooru Taniguchi and Gary P. Morriss}

\affiliation{School of Physics, University of New South Wales, Sydney, 
New South Wales 2052, Australia}

\date{\today}

\begin{abstract}

   Boundary effects in the stepwise structure of the Lyapunov spectra 
and the corresponding wavelike structure of the Lyapunov vectors are 
discussed numerically in quasi-one-dimensional systems consisting 
of many hard-disks. 
   Four kinds of boundary conditions constructed by combinations of 
periodic boundary conditions and hard-wall boundary conditions 
are considered, and lead to different stepwise structures of the 
Lyapunov spectra in each case. 
   We show that a spatial wavelike structure with a time-oscillation 
appears in the spatial part of the Lyapunov vectors divided by momenta 
in some steps of the Lyapunov spectra, while a rather stationary 
wavelike structure appears in the purely spatial part of the Lyapunov 
vectors corresponding to the other steps.  
   Using these two kinds of wavelike structure we categorize 
the sequence and the kinds of steps of the 
Lyapunov spectra in the four different boundary condition cases. 

\vspace{1cm}
\end{abstract}


\maketitle


\section{Introduction}

   Chaos is one of the essential concepts to justify 
a statistical treatment of deterministic dynamical systems.
   In a chaotic system a small initial error diverges 
exponentially, as characterized quantitatively 
by the Lyapunov exponents $\lambda_{n}$, 
and it means that it is impossible in principle to 
predict precisely any quantity of deterministic systems 
at any time and a statistical treatment of the 
systems is required. 
   It is well known that even one-particle systems can be chaotic 
and have some of the important statistical properties of 
equilibrium statistical mechanics such as the mixing property, etc. 
   For this reason many works on the subject of chaos have been 
done in one-particle systems, for example, billiard systems and 
Lorentz gas models \cite{Gas98,Dor99}. 
   However many-particle effects should still play an important 
role in some statistical aspects, such as the central limit theorem and 
a justification of thermodynamical reservoirs, etc. 
   Therefore it should be interesting to know what happens with
the combination of a chaotic effect and a many-particle effect, 
in other words, to know which chaotic effects do not 
appear in one-particle systems. 

   The Lyapunov spectrum is introduced as 
the sorted set $\{\lambda_{1}, \lambda_{2}, \cdots \}$ of 
Lyapunov exponents satisfying the condition 
$\lambda_{1} \geq \lambda_{2} \geq \cdots$, and 
is a quantity to 
characterize the many-particle chaotic dynamics.
   Recently a stepwise structure of the Lyapunov spectrum was 
reported numerically, as one of the many-particle chaotic effects, 
in many-disk systems in which each disk interacts with the 
other disks by hard-core interactions \cite{Del96,Mil98a,Pos00}. 
   This stepwise structure appears in 
the region of small absolute values of Lyapunov exponents. 
   This fact suggests that steps in the Lyapunov spectra 
come from slow and macroscopic behavior  
of the system, because small positive Lyapunov exponents should 
correspond to slow relaxation precesses. 
   This point is partly supported by the existence of a global  
structure in the Lyapunov vectors, the so called Lyapunov modes, 
which are wavelike structures in the tangent space of 
each eigenvector of a degenerate Lyapunov exponent, that is, 
for each stepwise structure \cite{Mil98a,Pos00,Mcn01a,Pos02a}. 
   The wavelike structure of the Lyapunov vectors appears 
as a function of the particle position, so this structure is also 
important as a relation to connect the tangent space with the 
phase space. 
   Although the Lyapunov vectors have been the subject 
of some works more than a decade ago 
(for example, see Refs. 
\cite{Kan86,Ike86,Gol87,Kon92,Fal91,Yam98,Pik01,Yam02}), 
it is remarkable that  
an observation of their global structure in fully chaotic 
many-particle systems has only recently appeared. 
   Explanations for the stepwise structure of the Lyapunov spectra 
have been attempted using periodic orbit models \cite{Tan01} and 
using the master equation approach \cite{Tan02}.  
   Other approaches to the Lyapunov modes have included using a random 
matrix approach for a one-dimensional model \cite{Eck00} and 
using a kinetic theoretical approach \cite{Mcn01b}, 
by considering these as the "Goldstone modes" \cite{Wij02}. 

   If the stepwise structure of Lyapunov spectra is a reflection 
of a global behavior of the system, then one may ask the question: 
Does such a structure depend on boundary conditions 
which specify the global structure of the system?  
   One of the purposes of this paper is to answer this question 
using some simple examples. 
   In this paper we investigate numerically the stepwise structure 
of the Lyapunov spectra in many-hard-disk systems  
of two-dimensional rectangular shape with 
the four kinds of the boundary conditions: 
[\textit{A1}] purely periodic boundary conditions, 
[\textit{A2}] periodic boundary conditions in the $x$-direction and 
hard-wall boundary conditions in the $y$-direction, 
[\textit{A3}] hard-wall boundary conditions in the $x$-direction and 
periodic boundary conditions in the $y$-direction, and 
[\textit{A4}] purely hard-wall boundary conditions. 
In all cases we took the $y$-direction as the narrow direction of 
the rectangular shape of the system and $x$-direction as the longer
orthogonal direction.   
   The case [\textit{A1}] means that the global shape of the system 
is like the surface of a doughnut shape, 
the cases [\textit{A2}] and [\textit{A3}] means 
that the system has the shape of the surface of pipe 
with hard-walls in its ends.
Case [\textit{A2}] is a long pipe with small diameter while case 
[\textit{A3}] is a short pipe with a large diameter.
   The case [\textit{A4}] means that the system has just a rectangular 
shape surrounded by hard-walls. 
   Adopting hard-wall boundary conditions in a particular direction
destroys the total momentum conservation in that direction, so by 
considering these models we can investigate the effects 
of the total momentum conservation on the stepwise structure 
of the Lyapunov spectra and the Lyapunov modes. 
   This can be used to check some theoretical approaches 
to these phenomena like those in Refs. \cite{Tan02,Eck00,Mcn01b}, 
in which the total momentum conservation plays an essential role 
in explaining the stepwise structure of the Lyapunov spectra or 
the Lyapunov modes. 
   We obtain different stepwise structures of the 
Lyapunov spectra with each different boundary condition case. 
   Especially, we observe a stepwise structure of the Lyapunov spectrum 
even in the purely hard-wall boundary case [\textit{A4}], in which the 
total momentum is not conserved in any direction. 

   The second purpose of this paper is to categorize 
the stepwise structure of the Lyapunov spectra according to the
wavelike structure of the Lyapunov vectors. 
   So far wavelike structure of the Lyapunov vectors was 
reported in the Lyapunov vector components as a function 
of positions, for example in the quantity $\delta q_{yj}^{(n)}$ 
as a function of the position $q_{xj}$ (the transverse Lyapunov mode) 
\cite{Mil98a,Pos00,Mcn01a} and in the quantity $\delta q_{xj}^{(n)}$ 
as a function of the position $q_{xj}$ 
(the longitudinal Lyapunov mode) \cite{Pos02}, 
in which $\delta q_{yj}^{(n)}$ 
($\delta q_{xj}^{(n)}$) is the $y$-component ($x$-component) 
of the spatial coordinate part of the 
Lyapunov vector of the $j$-th particle corresponding to 
the $n$-th Lyapunov exponent $\lambda_{n}$, and 
$q_{xj}$ is the $x$-component of the spatial component of the 
$j$-th particle. 
   These wavelike structures appear in the stepwise region of 
the Lyapunov spectrum. 
   However it is not clear whether there is a direct connection 
between the sequence and the kinds of steps of the Lyapunov 
spectrum and the modes of wavelike structure of the Lyapunov vectors 
in the numerical studies, 
that is, how we can categorize the steps of the Lyapunov spectrum 
by the Lyapunov modes. 
   In this paper we show another type of wavelike 
structure of the Lyapunov vectors, which appears in the 
quantity $\delta q_{yj}^{(n)}/p_{yj}$ with the $y$-component 
$p_{yj}$ of the momentum coordinate of the $j$-th particle, 
and use this wavelike structure to categorize 
the stepwise structure of the Lyapunov spectra.  
   The Lyapunov vector components $\delta q_{xj}^{(n)}$, 
$\delta q_{yj}^{(n)}$ and $\delta q_{yj}^{(n)}/p_{yj}$ 
have a common feature in which each quantity gives 
a constant corresponding to one of the zero Lyapunov exponents of the system 
with purely periodic boundary conditions, because of 
the conservation of center of mass or the deterministic nature of the orbit. 
   More concretely in this paper 
we consider the two kinds of graphs: 
[\textit{B1}] the quantity $\delta q_{yj}^{(n)}$ as a function of 
the position $q_{xj}$, namely the transverse Lyapunov mode, and 
[\textit{B2}] the quantity $\delta q_{yj}^{(n)}/p_{yj}$ as 
a function of the position $q_{xj}$ and time (or collision number). 
   In the two-dimensional rectangular systems consisting of 
many-hard-disks with periodic boundary conditions 
(the boundary case [\textit{A1}]) it is known that there are two types 
of steps of the Lyapunov spectra: steps consisting of two 
Lyapunov exponents and steps consisting of four Lyapunov 
exponents \cite{Pos00,Mcn01a}. 
   The wavelike structure in the graph [\textit{B1}] corresponding to 
the two-point steps of the Lyapunov spectrum for such a rectangular 
system is well known. 
   In this paper we show the wavelike structure corresponding 
to the four-point steps of the Lyapunov spectrum 
in the graph [\textit{B2}]. 
   Besides, we observe time-dependent oscillations in 
the graph [\textit{B2}],  
whereas the graph [\textit{B1}] is rather stationary in time. 
   The wavelike structure in the graph [\textit{B2}] also appears in the 
rectangular systems with a hard-wall boundary condition 
(the boundary cases [\textit{A2}], [\textit{A3}] and [\textit{A4}]),  
   specifically even in the case 
of purely hard-wall boundary conditions 
in which the transverse Lyapunov mode [\textit{B1}] does not appear. 
   We show that the stepwise structure of the Lyapunov 
spectra in the boundary cases [\textit{A1}], [\textit{A2}], [\textit{A3}] 
and [\textit{A4}] 
can be categorized by the wavelike structures of the graphs 
[\textit{B1}] and [\textit{B2}]. 
   
    One of the problems which make it difficult to investigate 
the structure of the Lyapunov spectra and the Lyapunov modes is that 
the calculation of a full Lyapunov spectra for many-particle 
system is a very time-consuming numerical calculation. 
   Therefore it should be important to know  
how we can calculate the stepwise structure of the 
Lyapunov spectra and the Lyapunov modes with as little calculation 
time as possible. 
   It is known that a rectangular system has a wider  
stepwise region of the Lyapunov spectrum than a 
square system \cite{Pos00}. 
   Noting this point, in this paper 
we concentrate on the most strongly rectangular system, 
namely on the quasi-one-dimensional system in which the 
rectangle is too narrow to allow particles to exchange their 
positions. 
   As will be shown in this paper, the stepwise structure of the 
Lyapunov spectrum for the quasi-one-dimensional system 
is the same as the fully rectangular system which allows  
exchange of particle positions and each particle can collide 
with any other particle, and the steps of the Lyapunov spectra 
consist of two-point steps and four-point steps in the periodic 
boundary case [\textit{A1}]. 
   We also determine an appropriate particle density to give
clearest stepwise structure of the Lyapunov spectrum 
in the quasi-one-dimensional system. 

   The outline of this paper is as follows. 
   In Sect. \ref{QuasiOneDime} we discuss how we 
can get the stepwise structure of the Lyapunov spectra for a small 
number of particle systems with a quasi-one-dimensional shape. 
   The density dependence of the Lyapunov spectrum 
for the quasi-one-dimensional system is investigated to get
a clearly visible stepwise structure of the Lyapunov spectra. 
   In Sect. \ref{PerioPP} we consider the purely periodic 
boundary case (the boundary case [\textit{A1}]), and 
investigate wavelike structures in the graphs [\textit{B1}] 
and [\textit{B2}] of Lyapunov vector components. 
   In Sect. \ref{PpPwWpWW} we consider the cases 
including a hard-wall boundary condition (the boundary cases 
[\textit{A2}], [\textit{A3}] and 
[\textit{A4}]) with calculations of the graphs [\textit{B1}] 
and [\textit{B2}], and 
compare the stepwise structures of the 
Lyapunov spectra and the wavelike structures in the graphs 
[\textit{B1}] and [\textit{B2}] in the above four boundary cases. 
   Finally we give our conclusion and remarks in Sect. \ref{Concl}.

%

\section{
Quasi-one-dimensional systems  
and density dependence of the Lyapunov spectrum}
\label{QuasiOneDime}
 
   The stepwise structure of the Lyapunov spectra is purely 
a many-particle effect of the chaotic dynamics, and so far it 
has been investigated in systems of 100 or more  particles.
    However the numerical calculation of Lyapunov 
spectra for such large systems is not so easy even at present. 
   Noting this point, in this section we discuss briefly how we 
can investigate the stepwise structure of the Lyapunov spectrum 
for a system whose number of particles is as small as possible.  
   We also discuss what is a proper particle density to 
get clear steps of the Lyapunov spectrum.
   These discussions give reasons to choose some values of 
system parameters used in the following sections. 

\begin{figure}[!htb]
\vspfigA
\includegraphics[width=\widthfigA]{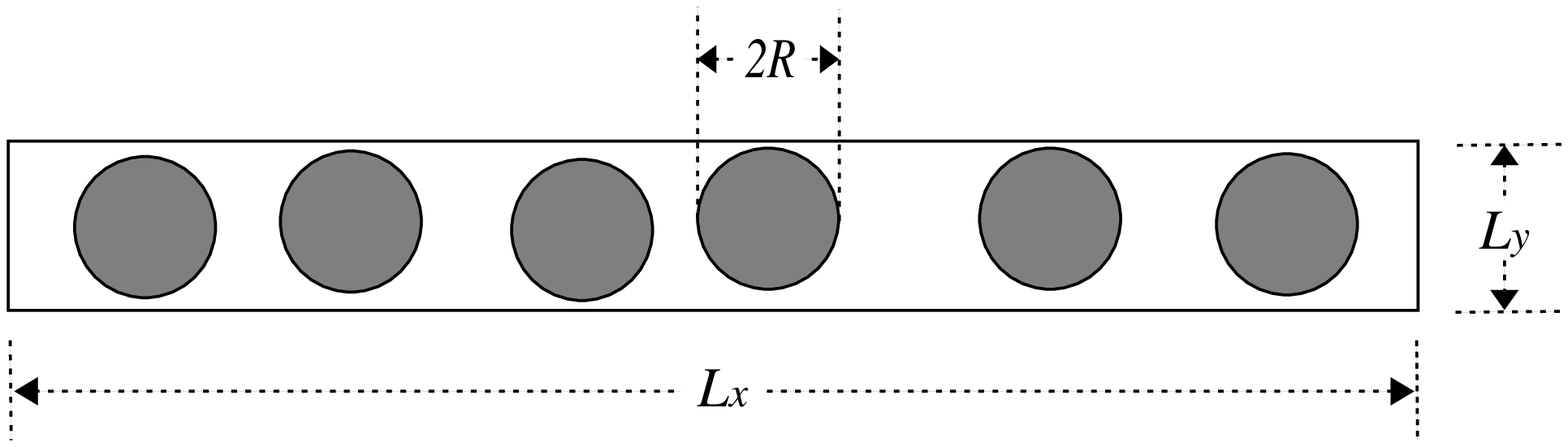}
\caption{Quasi-one-dimensional system: A narrow rectangular 
   system satisfying the conditions 
   $2RN < L_{x}$ and $2R < L_{y} < 4R$.}
\vspfigB
\label{qua1dim}\end{figure}  

   We consider two-dimensional systems consisting of $N$ number of 
hard-disks 
in which the radius of the particle is $R$ and the width (height) 
of the system is $L_{x}$ ($L_{y}$). 
   One way to get the stepwise structure of Lyapunov spectrum 
in a two-dimensional system consisting of a small number of particles 
is to choose a rectangular system rather than a square system, 
because the stepwise region in the Lyapunov spectrum is 
wider in a more rectangular system rather than in a square system 
even if the number of particles is the same \cite{Pos00}. 
   Noting this characteristic, we concentrate on 
the most rectangular case, namely 
the quasi-one-dimensional system defined by the conditions
   
\begin{eqnarray}
   2RN < L_{x} \;\; \mbox{and} \;\; 2R < L_{y} < 4R.    
\label{qua1dimcon}\end{eqnarray}

\begin{figure}[!htb]
\vspfigA
\includegraphics[width=\widthfigB]{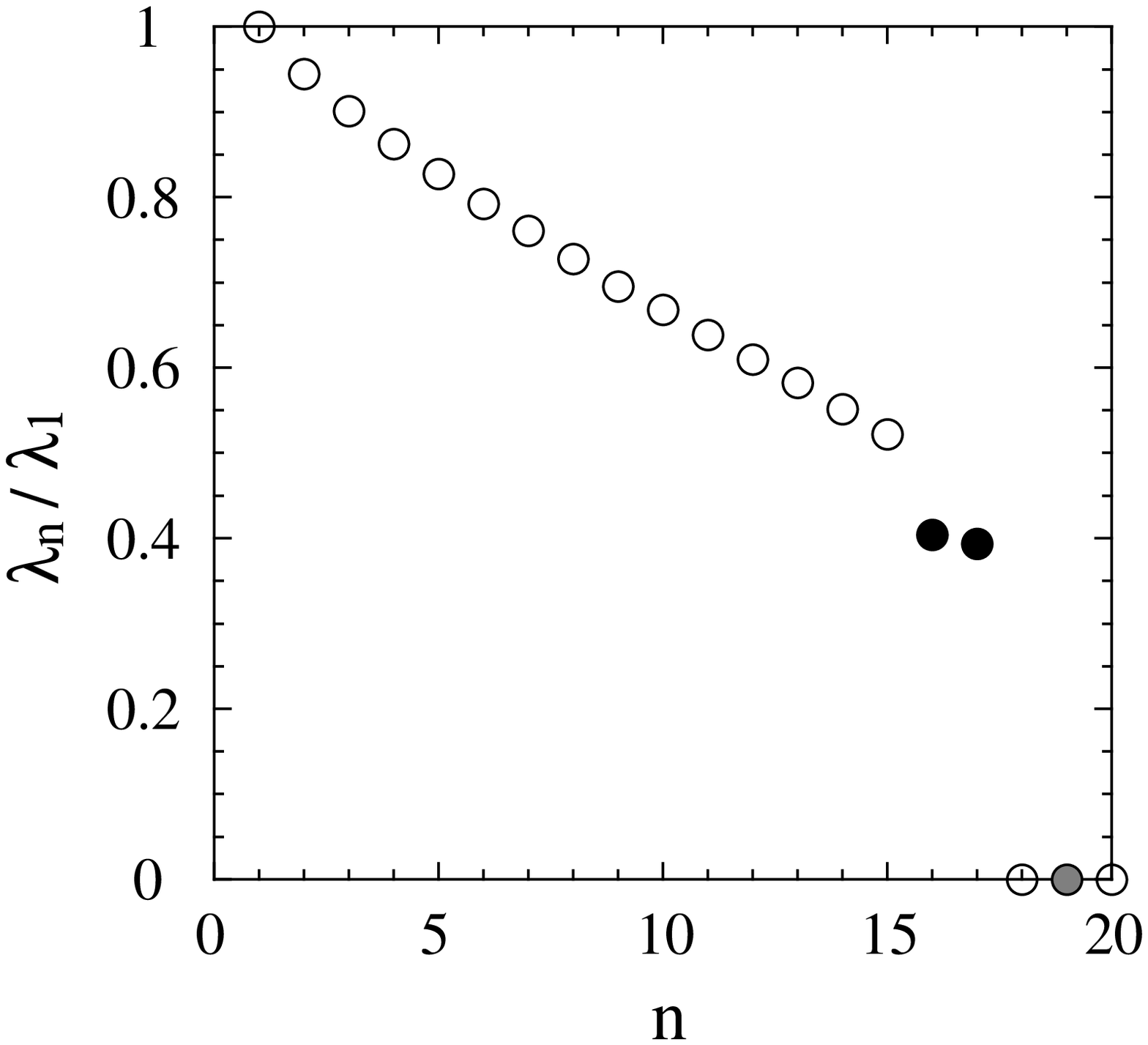}
\caption{
      Lyapunov spectrum normalized by the maximum Lyapunov 
   exponent in a 10-hard-disk system with 
   the periodic boundary conditions in both directions. 
      The Lyapunov exponents $\lambda_{16}$ and $\lambda_{17}$ 
   shown as the black circles form a two-point step 
   accompanying the transverse Lyapunov modes shown 
   in Fig. \ref{10mod}. 
      The gray circle is the zero-Lyapunov 
   exponent $\lambda_{19}$ whose corresponding 
   Lyapunov vector components $\delta q_{yj}^{(n)}$ 
   show a constant behavior (also see Fig.  \ref{10mod}).}
\vspfigB
\label{10lya}\end{figure}

\noindent A schematic illustration of the quasi-one-dimensional 
system is shown in Fig. \ref{qua1dim}. 
   In other words, the quasi-one-dimensional 
system is introduced as a narrow rectangular system where 
each particle can interact with two 
nearest-neighbor particles only and particles do not exchange 
the order of their positions. 
   In the quasi-one-dimensional system the upper bound 
$\rho_{max}$ of the 
particle density $\rho \equiv N \pi R^{2}/ L_{x}L_{y}$ 
is given by $\rho_{max} = \pi/4 = 0.7853\cdots$.
   In such a system we can get a stepwise structure of the 
Lyapunov spectrum even in a 10-particle system ($N=10$), as shown 
in Fig. \ref{10lya}, which is the Lyapunov 
spectrum normalized by the maximum Lyapunov exponent 
$\lambda_{1} \approx 3.51$. 
   To get this figure we chose the parameters as $R=1$, 
$L_{y} = 2R\times (1+10^{-6})$, $L_{x}=NL_{y}\times (1+10^{-3})$, 
and the mass $M$ of the particle and  the total energy 
$E$ are given by $1$ and $N$, respectively, 
and we used periodic boundary conditions in both the 
directions. 
   The particle density of this system is given by 
$\rho =     0.7846\cdots$. 
   Noting the pairing property of the Lyapunov spectrum 
for Hamiltonian systems, namely the property that in  
Hamiltonian systems any positive Lyapunov exponent accompanies 
a negative Lyapunov exponent with the same absolute value 
\cite{Arn89,Gas98},  
we plotted the first half of the Lyapunov spectrum in Fig. 
\ref{10lya}. 
   (The same omission of the negative branch 
of Lyapunov spectra from plots will be used throughout this paper.)
   It is clear that the Lyapunov exponents $\lambda_{16}$ 
and $\lambda_{17}$ form a two-point step in this 
Lyapunov spectrum. 

   In order to calculate Lyapunov spectra and Lyapunov vectors 
we use the algorithm due to Benettin et al., 
which is characterized by intermittent rescaling 
and renormalization of Lyapunov vectors \cite{Ben76,Shi79}. 
   In the application of this algorithm 
to systems with hard-core particle 
interactions, we calculate the matrix $\Lambda(t_{k})$, 
whose column vectors give the Lyapunov vectors  
$\delta\mathbf{\Gamma}_{n}(t_{k})$ corresponding to 
the local-time Lyapunov exponent $\tilde{\lambda}_{n}(t_{k})$  
at time $t=t_{k}$ just after the $k$-th collision 
of particles. 
   The dynamics of the matrix $\Lambda(t_{k})$ is 
given by $\Lambda(t_{k+1}) =  \mathcal{N}_{k} 
\mathcal{G}_{k} \mathcal{M}_{k}\Lambda(t_{k})$, in which 
$\mathcal{N}_{k}$ is the matrix required to normalize each column vector 
of the operated matrix,  $\mathcal{G}_{k}$ is 
the Gram-Schmidt procedure  ensuring 
the orthogonality of the columns of the operated matrix, and  
$\mathcal{M}^{(k)}$ specifies the tangent space dynamics including 
a free flight and a particle collision  \cite{Del96}. 
   The local-time Lyapunov exponent $\tilde{\lambda}_{n}(t_{k})$ 
is calculated as the exponents of the exponential 
divergence (contraction) from the $n$-th column vectors of the matrix 
$\Lambda(t_{k})$ to the $n$-th column vectors of the matrix  
$\mathcal{G}_{k} \mathcal{M}_{k}\Lambda(t_{k})$. 
   The Lyapunov exponent $\lambda_{n}$ is given as 
a time-averaged local-time Lyapunov exponent 
after a long time calculation: $\lambda_{n} = 
\lim_{k\rightarrow\infty}\tilde{\lambda}_{n}(t_{k})$. 
   Here we use the standard metric 
$d\delta s^{2} = \sum_{j=1}^{N} [ 
d \delta q_{xj}^{(n)}{}^{2} + d \delta q_{yj}^{(n)}{}^{2}
+ d \delta p_{xj}^{(n)}{}^{2} + d \delta p_{yj}^{(n)}{}^{2} 
]$ for the tangent space 
with the $x$-component $\delta p_{xj}^{(n)}$ 
and $y$-component $\delta p_{yj}^{(n)}$ of the Lyapunov vector 
of the $j$-th particle corresponding to the local-time 
Lyapunov exponent $\tilde{\lambda}_{n}$.  
   Other articles such as Refs. \cite{Ben76,Shi79,Del96} 
should be referred to for more details of the Benettin algorithm 
and the tangent space dynamics of many-hard-disk systems. 

   It is very important to note that there are two ways 
that convince us of the structure of the Lyapunov spectrum; 
one is simply to find a step structure directly in the Lyapunov spectrum, 
and another is to find a structure in the Lyapunov vectors corresponding 
to specific Lyapunov exponents.  
   Fig. \ref{10mod} is the plot of the time-average 
$\langle \delta q_{yj}^{(n)} \rangle$ of 
the $y$-component  of the $j$-th  
particle of the spatial part  
of the Lyapunov vector $\delta\mathbf{\Gamma}_{n}$ 
corresponding to the Lyapunov exponents $\lambda_{n}$, for
$n=19, 17$ and $16$ 
as functions of the time-average 
$\langle q_{xj} \rangle/L_{x}$ of the normalized 
$x$-component of 
spatial coordinate  of the $j$-th particle 
($j=1,2,\cdots,N$) (the graph [\textit{B1}]). 
   The corresponding Lyapunov exponents 
$\lambda_{17}$ and $\lambda_{16}$ are shown 
as the black-filled circles in Fig. \ref{10lya} 
and construct a two-point step of the Lyapunov spectrum. 
   To take the time average of the quantities $\delta q_{yj}^{(n)}$ 
and $q_{xj}$ we picked up these values just after particle 
collisions over $100N$ collisions and take the arithmetic 
averages of them. 
   The step consisting of two points in the Lyapunov spectrum of Fig. 
\ref{10lya} accompanies wavelike structures in their 
Lyapunov vectors, which are called the transverse Lyapunov modes.
   It should be emphasized that the Lyapunov modes in 
Fig. \ref{10mod} are rather stationary over $100N$ 
particle collisions and to take the time average can be 
a useful way to get clear their wavelike structures. 
   In this figure we fitted the numerical data for the Lyapunov modes 
corresponding to the Lyapunov exponents $\lambda_{n}$, 
$n=17$ and $16$, to the function  
$y=\alpha_{n}\cos(2\pi x + \beta_{n})$ 
($(\alpha_{17},\beta_{17}) = (0.19913,0.9703)$ 
for the triangle dots, and 
$(\alpha_{16},\beta_{16}) = (0.24025,-0.59677)$ 
for the square dots). 
   It should be noted that the difference $\beta_{17} - 
\beta_{16} = 0.9703 - (-0.59677) 
= 1.56707$ of the two values of the phases $\beta_{n}$ is approximately 
$\pi/2 = 1.570\cdots$, meaning that these two waves are 
orthogonal with each other. 
   The graph of the Lyapunov vector component 
$\delta q_{yj}^{(n)}$ as a function of the position $q_{xj}$ 
($j=1,2,\cdots,N$) become constant in one of the zero Lyapunov 
exponent $\lambda_{19}$ shown as 
the gray-filled circle in Fig. \ref{10lya}.
   In Fig. \ref{10mod} we also fitted the numerical data 
corresponding to the 
zero-Lyapunov exponent $\lambda_{19}$ by the constant 
function $y=\alpha_{19}$ with the fitting parameter value 
$\alpha_{19}=0.015114$.

\begin{figure}[!htb]
\vspfigA
\caption{
      Time-averaged $y$-components 
   $\langle \delta q_{yj}^{(n)} \rangle$ 
   of the spatial part of the Lyapunov 
   vector of the $j$-th particle 
   as functions of the time-average $\langle q_{xj} \rangle/L_{x}$ 
   of the normalized $x$-component 
   of spatial coordinate of the $j$-th particle 
   corresponding to the Lyapunov exponents $\lambda_{n}$, 
   $n=19, 17$ and $16$, in the 10-hard-disk system with 
   the periodic boundary conditions in both directions. 
      The circle, triangle and square dots correspond to 
   the Lyapunov exponents $\lambda_{19}$, $\lambda_{17}$ and 
   $\lambda_{16}$, respectively, 
   which are shown as   
   the black- and gray-filled circles in Fig. \ref{10lya}.
      The dotted and broken lines are the fitting lines 
   for the sinusoidal functions, and the solid line is the 
   fitting for the constant function.  
   }
\vspfigB
\label{10mod}\end{figure}  

   Although we can recognize the two-point step of the Lyapunov 
spectrum in Fig. \ref{10lya}, it is important to note that 
there is another 
type of steps in the two-dimensional hard-disk system with a 
rectangular shape and periodic boundary conditions. 
   It is well known that in such a system the Lyapunov spectrum can 
have a stepwise structure consisting of two-point steps and 
four-point steps \cite{Pos00,Mcn01b}. 
   If we want to investigate the four-point steps of the 
Lyapunov spectrum we may have to consider a system consisting of 
more than 10 particles. 

   Now we mention an angle in components of the Lyapunov vectors. 
   Fig. \ref{10angDqDp} is the graph for the time-average 
$\langle \theta_{n} \rangle$
of the angle $\theta_{n}$ 
between the spatial part $\delta \mathbf{q}^{(n)}$ 
and the momentum part $\delta \mathbf{p}^{(n)}$ of the 
Lyapunov vector 
$\delta\mathbf{\Gamma}_{n} = (\delta \mathbf{q}^{(n)}, 
\delta \mathbf{p}^{(n)})^{T}$ where $^{T}$  is the transverse operation,  
and $\theta_{n}$ is defined by 

\begin{eqnarray}
   \theta_{n}\equiv 
		 \cos^{-1}\left(\frac{
		 \delta \mathbf{q}^{(n)} \cdot \delta \mathbf{p}^{(n)} }
		 {|\delta \mathbf{q}^{(n)}||\delta \mathbf{p}^{(n)}|
		 }\right) , 
\end{eqnarray}

\noindent except for the ones corresponding to the zero Lyapunov 
exponents. 
   Here in order to take the time-average of the angle $\theta_{n}$ 
we picked up its values just after particle collisions over $1000N$ 
collisions and took the arithmetic average of them. 
   This graph shows that the spatial part $\delta \mathbf{q}^{(n)}$ 
and the momentum part $\delta \mathbf{p}^{(n)}$ of the 
Lyapunov vector are pointed toward almost the same direction, 
especially for the Lyapunov vectors corresponding 
to small absolute values of Lyapunov exponents \cite{Mcn01b}. 
   This fact suggests that if we get a structure in the vector 
$\delta \mathbf{q}^{(n)}$ then we may get the similar structure 
in the vector $\delta \mathbf{p}^{(n)}$, and vice versa. 
   It should also be noted that this gives a justification 
for some approaches to Lyapunov exponents in which the 
Lyapunov exponents are calculated through the spatial 
coordinate part only (or the momentum part only) of 
the Lyapunov vector \cite{Bei98,Tan01}. 
   We may also note that the graph for $\theta_{n}$, 
$n=1,2,\cdots,2N-3$ 
and $\theta_{4N-n+1}$, $n=1,2,\cdots,2N-3$ are symmetric 
with respect to 
the line $y=1/2$, which comes from the symplectic property 
of the Hamiltonian dynamics. 

\begin{figure}[!htb]
\vspfigA
\caption{
   Time-averaged angle $\langle \theta_{n} \rangle$
   between the spatial part 
   $\delta \mathbf{q}^{(n)}$ and the momentum part 
   $\delta \mathbf{p}^{(n)}$ of the Lyapunov vector 
   $\delta\mathbf{\Gamma}_{n} = (\delta \mathbf{q}^{(n)}, 
   \delta \mathbf{p}^{(n)} )^{T}$ corresponding to the 
   Lyapunov exponent $\lambda_{n}$ 
   in the 10-hard-disk system with 
   the periodic boundary conditions in both directions. }
\vspfigB
\label{10angDqDp}\end{figure}  

   Another important point to get a clear stepwise structure 
for the Lyapunov spectrum in a small system is 
to choose a proper particle density of the system.  
   Even if we restrict our consideration to quasi-one-dimensional 
systems, the shape of the Lyapunov spectrum depends on the particle 
density, so we should choose a density that gives a clearly   
visible stepwise structure in the Lyapunov spectrum.  
   Fig. \ref{ppLya50d}(a) is the Lyapunov spectra 
normalized by the maximum Lyapunov exponents for 
quasi-one-dimensional systems consisting of 50 hard-disks ($N=50$) 
with periodic boundary conditions in both directions. 
   We also give an 
enlarged figure \ref{ppLya50d}(b) for the small Lyapunov 
exponent region.  
   Here the system parameters are given by $R=1$, 
$L_{y} = 2R(1+10^{-6})$, $L_{x} = NL_{y}(1+d)$, and we 
used the case of $M=1$ and $E=N$. 
   The five Lyapunov spectra correspond to the case of 
$d=10^{-4}$ (the circle dots, 
   the density $\rho = 0.7853\cdots$), 
the case of $d=10^{-1}$ (the triangle dots, 
   the density $\rho = 0.7139\cdots$), 
the case of $d=1$       (the square dots, 
   the density $\rho = 0.3926\cdots$), 
the case of $d=10$  (the diamond dots, 
   the density $\rho = 0.07139\cdots$), 
the case of $d=10^{2}$  (the upside-down triangle dots, 
the density $\rho = 0.007776\cdots$). 
   The maximum Lyapunov exponents $\lambda_{1}$ are 
given approximately by $3.62$, $2.44$, $0.934$, $0.279$
and $0.0579$ in the cases of 
$d=10^{-4}, 10^{-1}, 1, 10$ and $10^{2}$, respectively.
   As shown in Figs. \ref{ppLya50d}(a) and (b), 
for smaller values of the quantity 
$d$ (namely in the higher particle density) the gaps 
between the nearest steps 
of the Lyapunov spectrum become larger, although the 
stepwise region of the Lyapunov spectrum does not seem to 
depend on the quantity $d$. 
  This means that we can get a clear stepwise structure of the 
Lyapunov spectrum in the small $d$ case (namely at high density). 


\begin{figure}[!htb]
\vspfigA
\caption{
      Density dependence of Lyapunov spectrum normalized by the 
   maximum Lyapunov exponent for quasi-one-dimensional systems 
   consisting of 50 particles 
   with the periodic boundary conditions in both directions. 
      The five Lyapunov spectra correspond to the case of 
   $L_{y} = 2R(1+10^{-6})$, $L_{x} = NL_{y}(1+d)$ with 
   $d=10^{-4}$ (the circle dots), 
   the case of $d=10^{-1}$ (the triangle dots), 
   the case of $d=1$       (the square dots), 
   the case of $d=10$  (the diamond dots), 
   the case of $d=10^{2}$  (the upside-down triangle dots). 
      (a) Full scale. 
	  (b) A small positive Lyapunov exponent region 
   including the stepwise structure of the Lyapunov spectra. 
      In the figure (b) we filled dots by black (gray) 
   for the Lyapunov exponents whose corresponding 
   time-averaged Lyapunov vector components 
   $\langle \delta q_{yj}^{(n)} \rangle$ 
   of the spatial part of the Lyapunov 
   vector of the $j$-th particle 
   as functions of the time-averaged $x$-component 
   $\langle q_{xj} \rangle$ 
   of spatial coordinate of the $j$-th particle 
   show wavelike structures (constant behaviors).}
\vspfigB
\label{ppLya50d}\end{figure}  

   In Fig. \ref{ppLya50d}(b) the Lyapunov exponents 
accompanying wavelike structures (constant behaviors) 
in the time-averaged 
Lyapunov vector components $\langle \delta q_{yj}^{(n)}\rangle$,  
as functions of the position $\langle q_{xj}\rangle$, 
are shown as the dots blacked (grayed) out. 
   In the small $d$ case, looking from the zero Lyapunov exponents, 
the two-point step appears first 
(see the cases of $d=10^{-4},10^{-1}$ and $1$ 
in Fig. \ref{ppLya50d}(b)), 
whereas the four-point step appears first in the large $d$ case 
(see the cases of $d=10^{1}$ and $d=10^{2}$ in Fig. \ref{ppLya50d}(b)). 
   Besides, at least in the small $d$ case, the two-point steps and 
the four-point steps do not appear repeatedly (see the cases 
of $d=10^{-4}$ and $10^{-1}$ in Fig. \ref{ppLya50d}(b)). 
   These facts mean that the sequence of steps in the 
Lyapunov spectrum depends on the quantity $d$, namely the 
particle density. 

   Wavelike structures in the time-averaged Lyapunov 
vector components $\langle \delta q_{yj}^{(n)} \rangle$ 
as functions of the position $\langle q_{xj} \rangle$, 
namely the transverse Lyapunov modes,  
appear mainly in two-point steps of the Lyapunov spectra. 
   Therefore we can use these wavelike structures 
to distinguish two-point steps from four-point steps of the 
Lyapunov spectra. 
   However such a distinguishability sometimes 
seems to fail  
in steps of the Lyapunov spectra near a region where the 
Lyapunov spectra is changing smoothly. 
   Actually the transverse Lyapunov modes may appear even in 
some four-point steps, if they are near such a smoothly changing 
Lyapunov spectrum region. 
   On the other hand, in such a region of the Lyapunov spectra, 
fluctuation of Lyapunov vectors is rather large, and the 
wavelike structure of the Lyapunov vectors become vague. 
   In Fig. \ref{ppLya50d}(b) we did not  
indicate by black-filled dots  
the Lyapunov exponents whose corresponding Lyapunov vector 
components $\langle \delta q_{yj}^{(n)} \rangle$ 
as functions of the position $\langle q_{xj} \rangle$ 
show wavelike-structures with such a level of vagueness. 

   Based on the discussions in this section, in the following 
two sections we consider only the case whose system parameters are 
given by $N=75$, $R=1$, $M=1$ and $E=N$. 
   The height and the width of the system are given by 
$L_{y} = 2R(1+10^{-6})$ and $L_{x}=1.5NL_{y}$ (the density $\rho = 
0.5235\cdots$) in the purely periodic boundary case 
considered in the next section.
   In this case, as will be shown in the next section, 
we can recognize at least 2 clear four-point steps,  
and the two-point steps and the four-point steps of 
the Lyapunov spectrum appear repeatedly in the first few steps 
in the Lyapunov spectrum for the system with 
the purely periodic boundary conditions.  
   We always take the time averaged quantities 
$\langle \delta q_{yj}^{(n)}\rangle$ 
and $\langle q_{xj}\rangle$ 
of the quantities $\delta q_{yj}^{(n)}$ 
and $q_{xj}$, respectively 
as the arithmetic average of their values taken in the times    
immediately after particle collisions, over $100N$ collisions. 
   We calculated more than $2 \times 10^{5}$ collisions 
($10^{6}$ collisions in some of the models) in order 
to get the Lyapunov spectra and the Lyapunov vectors in 
the models considered in this paper.


\section{Quasi-One-Dimensional Systems with Periodic Boundary 
Conditions}
\label{PerioPP}

\begin{figure}[!htb]
\vspfigA
\includegraphics[width=\widthfigA]{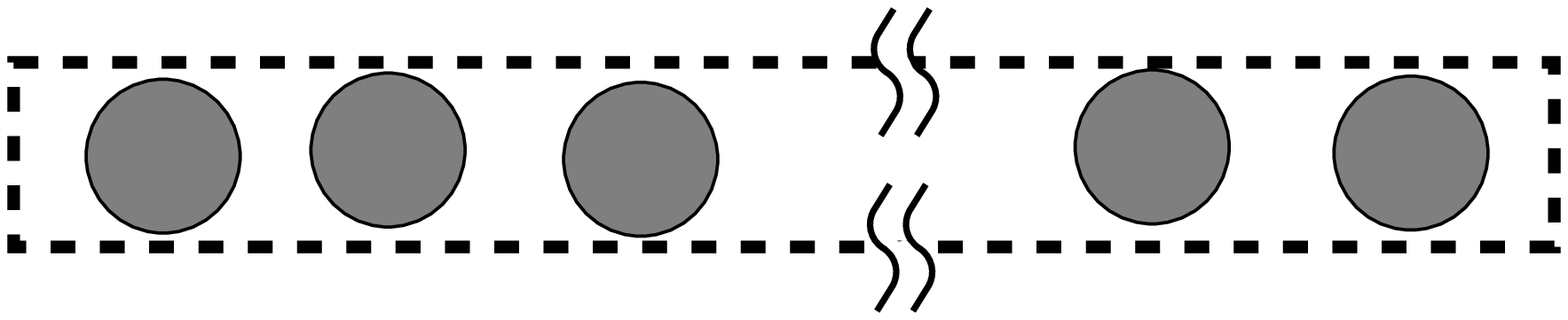}
\caption{A schematic illustration of a quasi-one-dimensional system 
   with periodic boundary conditions in both directions. 
   The broken lines indicate periodic boundary conditions 
   at that boundary.}
\vspfigB
\label{pp}\end{figure}

   In this section we consider the Lyapunov spectrum for the 
quasi-one-dimensional system with periodic boundary conditions 
in both the directions (the boundary case [\textit{A1}]). 
   A schematic illustration of this system for latter 
comparisons is given in Fig. \ref{pp} in which the broken line of 
the boundary means to take the periodic boundary conditions. 
   This system satisfies total momentum conservation in both 
the directions, and is regarded as a reference model for 
the models considered in the following section. 

   Fig. \ref{ppLya} is the small positive Lyapunov exponent region 
of the Lyapunov spectrum normalized by the 
maximum Lyapunov exponent $\lambda_{1}\approx 1.33$, 
including its stepwise region, for a 
quasi-one-dimensional system with periodic boundary conditions 
in both the directions. 
   The global shape of the Lyapunov spectrum is given in the 
inset in this figure. 
   We used the values of the system parameters chosen at the end 
of the preceding section. 
   At least 5 steps consisting 3 two-point steps and 
2 four-point steps are clearly visible in this Lyapunov spectrum. 

\begin{figure}[!htb]
\vspfigA
\caption{
      The stepwise structure of the Lyapunov spectrum 
   normalized by the maximum Lyapunov exponent 
   for a quasi-one-dimensional system 
   with periodic boundary conditions in both directions. 
      Inset: Full scale for the normalized Lyapunov spectrum. 
      The circular dots are filled by black (gray) 
   in the Lyapunov exponents corresponding to
   wavelike structures (a constant behavior) 
   of the time-averaged Lyapunov vector components  
   $\langle \delta q_{yj}^{(n)} \rangle$ shown in Fig. 
   \ref{ppDqy}.}
\vspfigB
\label{ppLya}\end{figure}  

   The two-point steps of the Lyapunov spectrum accompany 
wavelike structures in their corresponding Lyapunov vectors. 
   Fig. \ref{ppDqy} is the graphs of the time-averaged 
Lyapunov vector components $\langle \delta q_{yj}^{(n)} \rangle$ 
corresponding to the Lyapunov exponent $\lambda_{n}$, 
$n=149,147,146,141$ and $140$ as functions of 
the time-averaged position component 
$\langle q_{xj} \rangle$ 
normalized by the length $L_{x}$ 
(the graph [\textit{B1}]).  
   The Lyapunov exponents used for this figure are 
shown as the black or gray circles in Fig. \ref{ppLya}. 
   In this figure we also gave fittings of the numerical data 
by sinusoidal equations and a constant function.  
   The fitting equations are 
$y = \alpha_{n}\cos(2\pi x + \beta_{n})$ for $n=147,146$ 
( $(\alpha_{147},\beta_{147}) = (-0.16132,-4.5575), \;
(\alpha_{146},\beta_{146}) = (-0.16125,0.14928)$ ), 
and $y = \alpha_{n}\cos(4\pi x + \beta_{n})$ for $n=141,140$ 
( $(\alpha_{141},\beta_{141}) = (-0.15775,-0.30581), \; 
(\alpha_{140},\beta_{140}) = (0.15782,1.2666)$ ). 
   Concerning these fitting parameters,  
it is important to note the relations 
$|\alpha_{147}| \approx |\alpha_{146}|$, 
$|\alpha_{141}| \approx |\alpha_{140}|$,  
$\beta_{146}-\beta_{147} \approx 3\pi/2$ and 
$\beta_{140}-\beta_{141} 
\approx \pi/2$, meaning that the two wavelike structures in the 
same two-point step are orthogonal to each other. 
   In Fig. \ref{ppDqy} we also draw a graph of 
the averaged Lyapunov vector 
component $\langle \delta q_{yj}^{(n)} \rangle$ as a function 
of the normalized position $\langle q_{xj} \rangle/L_{x}$ 
corresponding to the 
second zero Lyapunov exponent $\lambda_{149}$, which is  
roughly speaking constant, and is fitted by a constant function 
$y=\alpha_{149}$ with the fitting parameter value  
$\alpha_{149} = 0.018901$. 
   These results suggest a conjecture for the approximate
form of the Lyapunov vector component $\delta q_{yj}^{(n(k))}$ 
corresponding to the Lyapunov exponents $\lambda_{n(k)}$ 
and $\lambda_{n(k)-1}$ in the same $k$-th two-point step 
counting from the zero Lyapunov exponents as 

\begin{eqnarray}
   &&
   \hspace{-0.5cm}
   \left\{\delta q_{yj}^{(n(k))}, \; \delta q_{yj}^{(n(k)-1)}
      \right\} \approx 
      \nonumber \\
	  \nonumber \\
   &&\left\{
       \alpha_{k}\cos\left(\frac{2\pi k}{L_{x}} q_{xj}
	   + \beta_{k} \right), \; 
	   \alpha_{k}\sin\left(\frac{2\pi k}{L_{x}} q_{xj} 
	   + \beta_{k} \right)
   \right\} ,   
       \nonumber \\ 
\label{LyapuWave}\end{eqnarray}

\noindent $j=1,2,\cdots,N$ with constants $\alpha_{k}$ 
and $\beta_{k}$. 
   (Note that in this paper we always count the sequence of  
steps of Lyapunov spectra looking from the zero Lyapunov 
exponents, so, for example, the two-point 
step consisting of the Lyapunov exponents $\lambda_{147}$ and 
$\lambda_{146}$ ($\lambda_{141}$ and 
$\lambda_{140}$) in the Lyapunov spectrum shown 
in Fig. \ref{ppDqy} is the first (second) two-point step.)
   It may be noted that the constant $\alpha_{k}$ 
in Eq. (\ref{LyapuWave}) should be 
determined by the normalization condition of the 
Lyapunov vector, which is used in the Benettin algorithm 
to calculate the Lyapunov spectrum and Lyapunov vectors 
in this paper. 

\begin{figure}[!htb]
\vspfigA
\caption{
   Time-averaged Lyapunov vector components 
   $\langle \delta q_{yj}^{(n)} \rangle$ as functions of the 
   time-average $\langle q_{xj} \rangle/L_{x}$ of 
   the normalized $x$-component
   of the spatial 
   coordinate of the $j$-th particle corresponding to the 
   $n$-th Lyapunov exponent $\lambda_{n}$ 
   for the quasi-one-dimensional system with the periodic 
   boundary conditions in both directions 
   ($n=149,147,146,141$ and $140$). 
      The corresponding Lyapunov exponents are shown as 
   the black- and gray-filled circles in Fig. \ref{ppLya}. 
      The numerical data are fitted by a constant function 
   and sinusoidal functions. 
   }
\vspfigB
\label{ppDqy}\end{figure}  

   It should be emphasized that the wavelike structures in Figs. 
\ref{10mod} and \ref{ppDqy} correspond to the two-point steps 
of the Lyapunov spectrum. 
   We cannot recognize such a clear wavelike structure in the graph 
of the Lyapunov vector component 
$\langle \delta q_{yj}^{(n)} \rangle$ 
as a function of the position $\langle q_{xj} \rangle$ 
corresponding to the four-point steps of the Lyapunov spectrum. 
   This suggests that the physical meaning of the four-point steps 
is different from the two-point steps, and we should consider 
a different quantity to investigate as a structure of Lyapunov vectors 
corresponding to the four-point steps of the Lyapunov spectrum. 
   Now, as one of the important results of this paper, we 
show that wavelike structures in the quantities 
$\delta q_{yj}^{(n)}/p_{yj}$
as functions of $q_{xj}$ and the collision number 
appear in the four-point steps of the Lyapunov spectrum. 
   As a motivation to introduce the quantity 
$\delta q_{yj}^{(n)}/p_{yj}$ to investigate its structure, 
we note that the small perturbation of the spatial 
coordinates of the particles in the direction of the orbit, 
namely the small perturbation 
$\delta \mathbf{q} \propto \mathbf{p}$, leads to 
a zero-Lyapunov exponent, so that the quantities 
$\delta q_{yj}^{(n)}/p_{yj}$, $j=1,2,\cdots,N$
should give a constant value corresponding to this zero Lyapunov exponent.  
   This is the common feature as 
the quantity $\delta q_{yj}^{(n)}$, which shows  
a constant behavior corresponding to one of the zero-Lyapunov 
exponents owing to the conservation of center of mass, 
and whose wavelike structure we have already discussed. 

   Before showing graphs of the quantities 
$\delta q_{yj}^{(n)}/p_{yj}$, 
we discuss some difficulties in the investigation of 
wavelike structure in these quantities. 
   It is much harder to get the wavelike structure of these 
quantities corresponding to the four-point steps of the 
Lyapunov spectrum compared to the quantities 
$\delta q_{yj}^{(n)}$ 
corresponding to the 
two-point steps of the Lyapunov spectrum, 
for at least two reasons. 
   As the first reason, fluctuation in the wavelike structure of  
the quantities $\delta q_{yj}^{(n)}/p_{yj}$ is 
much bigger than in the wavelike structure of  
the quantities $\delta q_{yj}^{(n)}$, partly because 
such a fluctuation is magnified in the case of a 
small absolute value of the quantity $p_{yj}$ appearing 
in the denominator of the quantity $\delta q_{yj}^{(n)}/p_{yj}$. 
   Secondly, the wavelike structure of the quantity 
$\delta q_{yj}^{(n)}/p_{yj}$, at least the magnitude 
of its wave, oscillates periodically in time, 
whereas the wavelike structure of the 
quantities $\delta q_{yj}^{(n)}$ in Fig. \ref{ppDqy} are 
stationary at least over more than $100N$ collisions. 
   This fact gives an upper bound on the time period  
(or the collision number interval) over which we can take the 
time-average of the quantities $\delta q_{yj}^{(n)}/p_{yj}$ 
in order to suppress the fluctuations 
and get their clear wavelike structures. 
   In this paper we express the local time-averages of the 
quantities $\delta q_{y}^{(n)}/p_{yj}$ and $q_{xj}$ as 
$\langle\delta q_{yj}^{(n)}/p_{yj}\rangle_{t}$ 
and $\langle q_{xj}\rangle_{t}$, respectively, 
with the suffix $t$ to remind us that they can change in time.

\begin{figure}[!htb]
\vspfigA
\caption{
      Local time-averaged quantities 
   $\langle\delta q_{yj}^{(n)}/p_{yj} \rangle_{t}$ 
   as functions of the normalized position 
   $\langle q_{xj} \rangle_{t}/L_{x}$ and the collision number 
   $n_{t}$ corresponding to (a) the Lyapunov 
   exponent $\lambda_{142}$ in the first four-point 
   step and (b) the Lyapunov exponent $\lambda_{139}$ in 
   the second four-point step
   in the same collision number interval $[543000,548100]$. 
      The system is the quasi-one-dimensional system with 
   the periodic boundary conditions in both directions, 
   and the corresponding Lyapunov exponents are 
   indicated by arrows in Fig. \ref{ppLya}. 
       In the contour plots on the bottoms of these 
   three-dimensional plots, red dotted lines, black solid lines and 
   blue broken lines correspond to 
   the values $\langle\delta q_{yj}^{(n)}/p_{yj} \rangle_{t} = 0.08, 
   0$ and $-0.08$, respectively.
    }
\vspfigB
\label{ppDqyPy}\end{figure} 

\begin{figure*}[!htb]
\vspfigA
\caption{
      Contour plots of the local-time averaged quantities  
   $\langle\delta q_{yj }^{(n)}/p_{yj}\rangle_{t}$ 
   as functions of  the normalized 
   position $\langle q_{xj}\rangle_{t}/L_{x}$ 
   and the collision number $n_{t}$ 
   corresponding to the first four-point step 
   consisting of the Lyapunov exponents $\lambda_{n}$, 
   $n=145,144,143$ and $142$ in the same collision number interval 
   $[538500,548100]$. 
      Here red dotted lines, black solid lines and 
   blue broken lines correspond to the values 
   $\langle\delta q_{yj}^{(n)}/p_{yj} \rangle_{t} = 0.08, 
   0$ and $-0.08$, respectively. 
      The system is the quasi-one-dimensional system with 
   the periodic boundary conditions in both directions, 
   and the corresponding Lyapunov exponents used in this figure 
   are indicated by the 
   brace under circles in Fig. \ref{ppLya}. }
\vspfigB
\label{ppDqyPyCont}\end{figure*}  

   In this and the next sections we will give the many graphs of 
$\delta q_{yj}^{(n)}/p_{yj}$ as functions of $q_{xj}$ 
and the collision number by taking their local time-averages, 
so here we summarize how we calculate the data for those graphs 
from a technical point of view.  
   First we take the arithmetic time-average 
$\langle\delta q_{yj}^{(n)}/p_{yj}\rangle_{t}$ and 
$\langle q_{xj}\rangle_{t}$ of the quantities 
$\delta q_{yj}^{(n)}/p_{yj}$ and $q_{xj}$, respectively, 
using their values just after particle 
collisions over $4N$ collisions 
($8N$ collisions), but if the absolute value $|p_{yj}|$ 
of the momentum is less than 10 percent 
(5 percent) of the averaged 
momentum amplitudes $\sqrt{2ME/N}$ then we exclude 
the quantity $\delta q_{yj}^{(n)}/p_{yj}$ at that time 
from samples to take this local time-average, 
in the models of this section and the subsection 
\ref{HarWalPW} (in the models of the subsections \ref{HarWalWP} 
and \ref{HarWalWW}). 
   (Therefore the sample number for taking the arithmetic 
averages can be less than $4N$ ($8N$) in the models of this section 
and the subsection \ref{HarWalPW}  
(in the models of the subsections \ref{HarWalWP} 
and \ref{HarWalWW}).) 
   On this local time-averages we can still get more than  
ten locally time-averaged datas for different times in one period 
of the time-oscillations for the graphs 
in the slowest time-oscillating movement of the quantity 
$\delta q_{yj}^{(n)}/p_{yj}$ corresponding to a step of 
the Lyapunov spectra, for example, corresponding to the 
first four-point step of the model in this section. 
   In this paper we consider the graph of the 
quantity $\langle\delta q_{yj}^{(n)}/p_{yj}\rangle_{t}$ 
as a function of $\langle q_{xj}\rangle_{t}$ and 
$n_{t}$ which is the first collision number of the 
collision number interval taking the local time averages 
$\langle \cdots \rangle_{t}$. 

   Figure \ref{ppDqyPy}(a) and (b) is the graphs of the quantity 
$\langle\delta q_{yj}^{(n)}/p_{yj}\rangle_{t}$ 
as functions of the normalized position 
$\langle q_{xj}\rangle_{t}/L_{x}$ 
and the collision number $n_{t}$ (the graph [\textit{B2}]), 
corresponding to the Lyapunov exponents 
$\lambda_{n}$, $n=145$ and $139$, respectively, 
indicated by arrows in Fig. \ref{ppLya}. 
   These two graphs correspond to Lyapunov exponents 
in different four-point steps and are given in the 
same collision number interval $[543000,548100]$. 
   In the graph corresponding to the Lyapunov exponent 
$\lambda_{145}$ ($\lambda_{139}$) we can recognize 
a spatial wavelike structure  
with the spatial wavelength $1$ ($1/2$) oscillating in time. 
   The time-oscillating period corresponding to 
the Lyapunov exponent $\lambda_{139}$ is about half of  
the period of the oscillation corresponding to 
the Lyapunov exponent $\lambda_{145}$. 
   These graphs, especially Fig. \ref{ppDqyPy}(b), are  
the most difficult graphs to recognize the structures  
from their three-dimensional plots in this paper,  
and in order to recognize the structures the contour plots 
given in the bottoms of these 
three-dimensional plots may be helpful. 
   In these contour plots we color-coded mountain regions 
($\langle\delta q_{yj}^{(n)}/p_{yj}\rangle_{t} > 0$) 
and valley regions 
($\langle\delta q_{yj}^{(n)}/p_{yj}\rangle_{t} > 0$) 
of the time-oscillating wavelike structure 
of the quantities   
$\langle\delta q_{yj}^{(n)}/p_{yj}\rangle_{t}$ by 
the red dotted contour lines of 
$\langle\delta q_{yj}^{(n)}/p_{yj}\rangle_{t}=0.08$ and 
the blue broken 
contour lines of 
$\langle\delta q_{yj}^{(n)}/p_{yj}\rangle_{t} = -0.08$, 
respectively, separating them by the black solid contour lines of 
$\langle\delta q_{yj}^{(n)}/p_{yj}\rangle_{t} = 0$.

\begin{figure}[!htb]
\vspfigA
\includegraphics[width=\widthfigC]{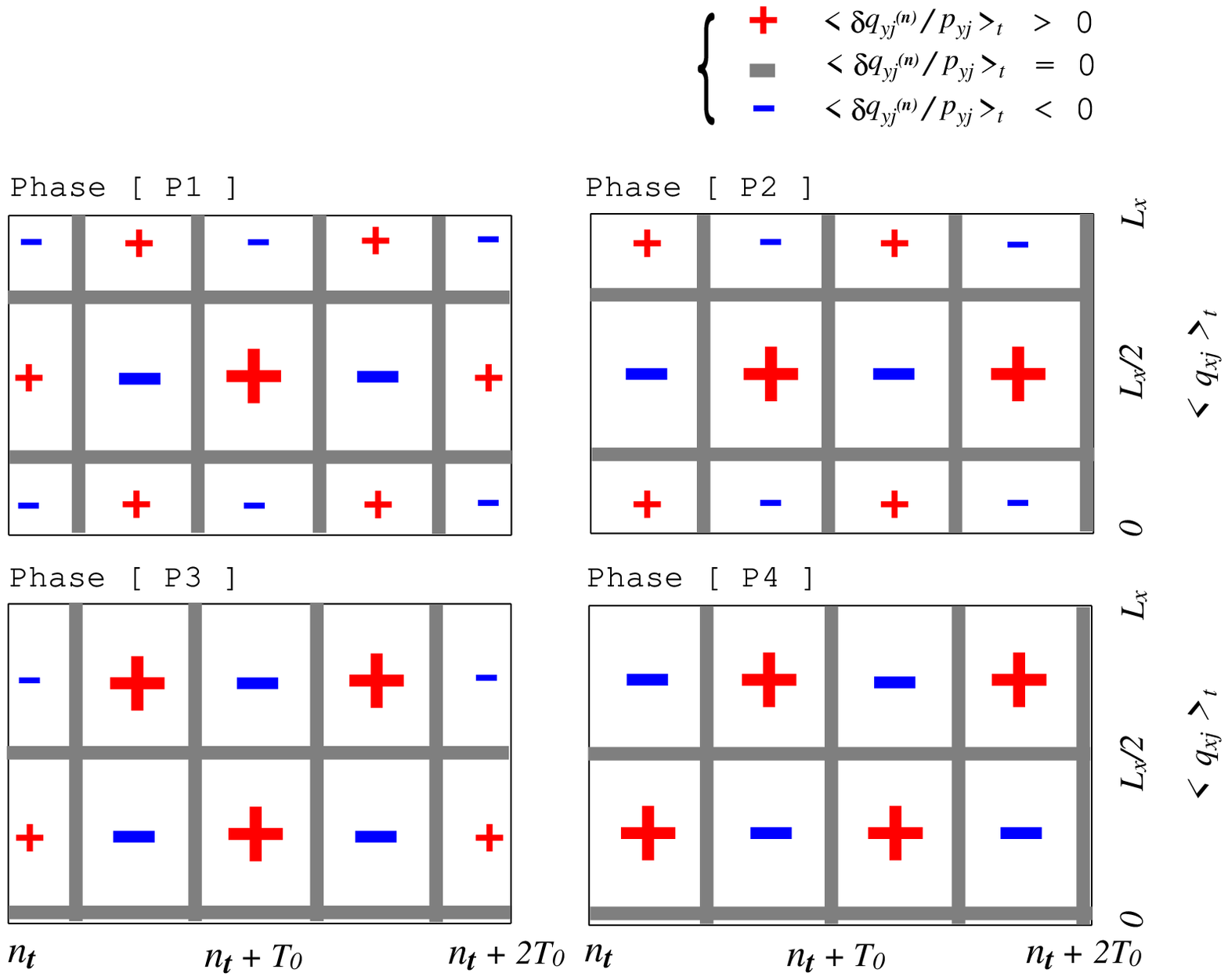}
\caption{
      Schematic illustration of the phase relations among 
   the four time-oscillating wavelike structures of the quantities  
   $\langle\delta q_{yj }^{(n)}/p_{yj}\rangle_{t}$ 
   as functions of the position $\langle q_{xj} \rangle_{t}$ 
   and the collision number $n_{t}$, corresponding 
   to the first four-point step in the same collision number interval. 
      The spatial wavelength and the period of the time-oscillations 
   are given by $L_{x}$ and $\mathcal{T}_{0}$, respectively.   
      Thick gray lines means node lines, 
   the region indicated by 
   a red plus sign ($+$) is the region where the quantity 
   $\langle\delta q_{yj }^{(n)}/p_{yj}\rangle_{t}$ is positive, 
   and the region indicated by 
   a blue minus sign ($-$) is the region where the quantity 
   $\langle\delta q_{yj }^{(n)}/p_{yj}\rangle_{t}$ is negative.
      The phases [\textit{P3}] and [\textit{P4}] differ from the
   phases [\textit{P1}] and [\textit{P2}] by a phase shift of
   $\pi/2$ in the spatial direction. 
   The phases [\textit{P1}] and [\textit{P3}] differ from the 
   phases [\textit{P2}] and [\textit{P4}] by a phase shift of 
   $\pi/2$ in time.
   }
\vspfigB
\label{phaseCont1}\end{figure}

   Now we consider a relation among the quantities  
$\langle\delta q_{yj}^{(n)}/p_{yj}\rangle_{t}$ 
corresponding to Lyapunov exponents in the same four-point 
step.  
   Fig. \ref{ppDqyPyCont} is the contour plots of the quantity 
$\langle\delta q_{yj }^{(n)}/p_{yj}\rangle_{t}$ 
as functions of the normalized position 
$\langle q_{xj}\rangle_{t}/L_{x}$  
and the collision number $n_{t}$ 
corresponding to the first four-point step 
consisting of the Lyapunov exponents $\lambda_{n}$, 
$n=145,144,143$ and $142$ in the same collision number 
interval $[538500,548100]$. 
   Here red dotted lines, black solid lines and 
blue broken lines correspond to 
$\langle\delta q_{yj}^{(n)}/p_{yj} \rangle_{t} = 0.08, 
0$ and $-0.08$, respectively. 
   It is clear that in these four graphs their spatial 
wavelengths (given by the system length $L_{x}$) and 
time-oscillating periods 
(given by a constant $\mathcal{T}_{0}$) are almost the same 
as each other. 
   On the other hand we can recognize that the position of the nodes 
of the spatial waves of the graphs \ref{ppDqyPyCont}(a) 
and \ref{ppDqyPyCont}(b), 
as well as the nodes of the spatial waves of the graphs 
\ref{ppDqyPyCont}(c) and \ref{ppDqyPyCont}(d), coincide 
with each other approximately, 
and the phase difference between the spatial 
waves of the graphs \ref{ppDqyPyCont}(a) and  
\ref{ppDqyPyCont}(c) is $\pi/2$ approximately.   
   Besides, the node of the time-oscillation of 
the wave-amplitude of the graphs \ref{ppDqyPyCont}(a) 
and \ref{ppDqyPyCont}(c), 
as well as the nodes of the time-oscillation of 
the wave-amplitude of the graphs 
\ref{ppDqyPyCont}(b) and \ref{ppDqyPyCont}(d), coincide 
with each other approximately, and the phase difference between 
time-oscillations of the graphs \ref{ppDqyPyCont}(a) 
\ref{ppDqyPyCont}(b) is about $\pi/2$. 
   These points are summarized in Fig. \ref{phaseCont1}, 
which is the schematic illustration 
of the phase relations among 
the time-oscillating wavelike structures of the quantity 
$\langle\delta q_{yj }^{(n)}/p_{yj}\rangle_{t}$, 
   Here the phase [\textit{P1}], [\textit{P2}], 
[\textit{P3}] and [\textit{P4}] correspond to 
Figs. \ref{ppDqyPyCont}(a), (b), (c) and (d), respectively. 
   These observations suggest that the Lyapunov vector 
components $\delta q_{yj}^{(\tilde{n}(k))}$, 
$\delta q_{yj}^{(\tilde{n}(k)-1)}$, 
$\delta q_{yj}^{(\tilde{n}(k)-2)}$ and 
$\delta q_{yj}^{(\tilde{n}(k)-3)}$ corresponding to the 
Lyapunov exponents constructing the $k$-th four-point step 
are approximately expressed as 

\begin{eqnarray}
   && \hspace{-0.8cm} 
   \left\{
      \delta q_{yj}^{(\tilde{n}(k))}, \;
      \delta q_{yj}^{(\tilde{n}(k)-1)}, \;
      \delta q_{yj}^{(\tilde{n}(k)-2)}, \;
      \delta q_{yj}^{(\tilde{n}(k)-3)} 
   \right\} \approx 
   \nonumber \\
   \nonumber \\
   && \hspace{-0.2cm} \left\{
      \tilde{\alpha}_{k} p_{yj}
	     \cos\left(\frac{2\pi k}{L_{x}} q_{xj}
		    \!+\! \tilde{\beta}_{k} \right) 
	     \cos\left( \frac{2\pi k}{\mathcal{T}_{0}} n_{t}
            \!+\! \tilde{\gamma}_{k} \right) ,\right.  
     \nonumber \\
   && \tilde{\alpha}_{k}p_{yj}
	     \cos\left(\frac{2\pi k}{L_{x}} q_{xj}
		    \!+\! \tilde{\beta}_{k} \right) 
	     \sin\left( \frac{2\pi k}{\mathcal{T}_{0}} n_{t}
            \!+\! \tilde{\gamma}_{k} \right)  ,
     \nonumber \\
   && \tilde{\alpha}_{k}p_{yj}
	     \sin\left(\frac{2\pi k}{L_{x}} q_{xj}
		    \!+\! \tilde{\beta}_{k} \right) 
	     \cos\left( \frac{2\pi k}{\mathcal{T}_{0}} n_{t}
            \!+\! \tilde{\gamma}_{k} \right)  ,
     \nonumber \\
   && \left. \tilde{\alpha}_{k}p_{yj}
	     \sin\left(\frac{2\pi k}{L_{x}} q_{xj}
		    \!+\! \tilde{\beta}_{k} \right) 
	     \sin\left( \frac{2\pi k}{\mathcal{T}_{0}} n_{t}
            \!+\! \tilde{\gamma}_{k} \right)  
   \right\} , \nonumber \\
\label{LyapuOscil}\end{eqnarray}

\noindent $j=1,2,\cdots,N$, with constants $\tilde{\alpha}_{k}$, 
$\tilde{\beta}_{k}$ and $\tilde{\gamma}_{k}$. 
   Here $\mathcal{T}_{0}$ is the period of the time-oscillation of 
the quantity $\delta q_{yj}^{(\tilde{n}(k))}/p_{yj}$ in 
the first four-point step. 
   (In this paper we use the quantity $\mathcal{T}_{0}$ as the 
period of the particle-particle collision number, but we can always 
convert it 
into the real time interval approximately by multiplying 
it with the mean free time, which is, for example, about $0.0248$ in 
the model of this section.)
   It may be emphasized that the level of approximation in Eq. 
(\ref{LyapuOscil}) for the four-point steps may be worse than in Eq. 
(\ref{LyapuWave}) for the two-point steps.

\section{Quasi-One-Dimensional Systems with 
a Hard-wall Boundary Condition
}
\label{PpPwWpWW}

   In this section we consider quasi-one-dimensional 
systems with boundary conditions including a hard-wall 
boundary condition. 
   Noting that there are two directions in which 
we have to introduce the boundary conditions, 
we consider the three cases of systems;  
[\textit{A2}] the case of periodic boundary conditions 
in the $x$-direction 
and hard-wall boundary conditions in the $y$-direction, 
[\textit{A3}] the case of hard-wall boundary conditions 
in the $x$-direction 
and periodic boundary conditions in the $y$-direction, and 
[\textit{A4}] the case of hard-wall boundary conditions in both the 
directions. 
   To adopt the hard-wall boundary condition means that 
there is no momentum conservation in the direction 
connected by hard-walls, 
so it allows us to discuss the role of momentum conservation 
in the stepwise structure of the Lyapunov spectrum by 
comparing models having a hard-wall boundary conditions 
with models having periodic boundary conditions. 
   We will get different stepwise structures for the Lyapunov 
spectra in the above three cases and the case of the preceding 
section, and the investigation of Lyapunov vectors 
corresponding to steps of the Lyapunov spectra leads 
to relating and to categorizing them with each other. 

   In the systems with a hard-wall boundary condition 
we should carefully choose the width $L_{x}$ and 
the height $L_{y}$ of the systems 
for meaningful comparisons between the results 
in the systems with different boundary conditions. 
   It should be noted that in systems with pure periodic 
boundary conditions, the centers of particles can reach
to the periodic boundaries, while in hard-wall 
boundary conditions the centers of particles can only 
reach within a distance $R$ (the particle radius) of
the hard-wall boundaries. 
   In this sense the effective region for particles to move 
in the system with hard-wall boundary conditions 
is smaller than in the corresponding system with periodic boundary 
conditions, if we choose the same lengths $L_{x}$ and $L_{y}$. 
   In this section the lengths  $L_{x}$ and $L_{y}$ 
of the systems with a hard-wall boundary condition are chosen so that the 
effective region for particle to move is the same as in 
the pure periodic boundary case considered in the previous section, 
and are given by 
$(L_{y},L_{x}) = (2R(1+10^{-6})+2R,1.5N(L_{y}-2R))$ 
in the case [\textit{A2}], 
$(L_{y},L_{x}) = (2R(1+10^{-6}),1.5NL_{y}+2R)$ 
in the case [\textit{A3}]  and 
$(L_{y},L_{x}) = (2R(1+10^{-6})+2R,1.5N(L_{y}-2R)+2R)$ 
in the case [\textit{A4}]. 
   These choices of the lengths  $L_{x}$ and $L_{y}$ also 
lead to the almost same mean free time in the four 
different boundary condition cases. 
   In the cases of [\textit{A2}] and [\textit{A4}] with 
this choice of the system width $L_{y}$, in principle 
there is a space to exchange 
the particle positions in the $x$-direction 
(namely, strictly speaking these cases do not satisfy the 
second condition of Eq. (\ref{qua1dimcon}) ), 
but the space is extremely narrow (that is $2R\times 10^{-6}$) 
so that it is almost impossible for particle positions to actually 
be exchanged.


\subsection{The case of periodic boundary conditions in the 
$x$-direction and hard-wall boundary conditions 
in the $y$-direction}
\label{HarWalPW}

\begin{figure}[!htb]
\vspfigA
\includegraphics[width=\widthfigA]{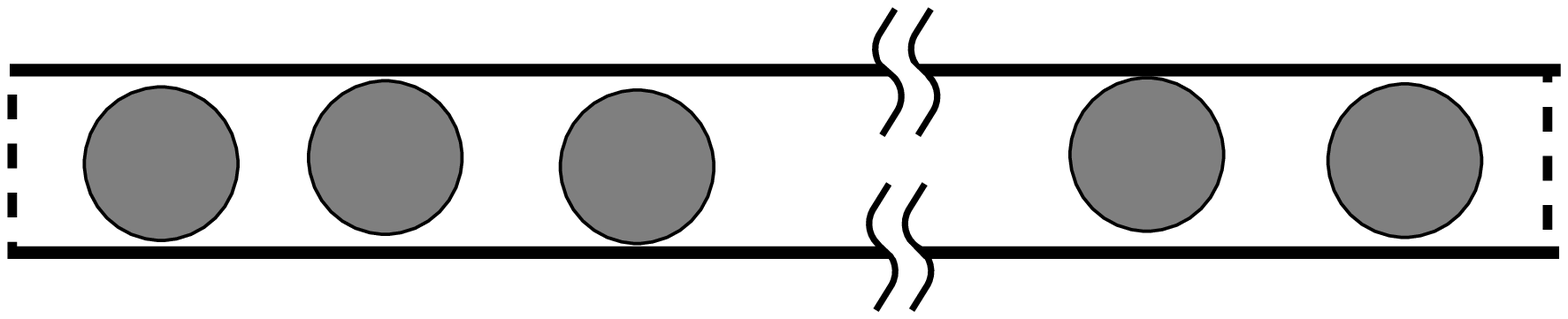}
\caption{
      A schematic illustration of a quasi-one-dimensional system 
   with periodic boundary conditions in the $x$-direction 
   and hard-wall boundary conditions in the $y$-direction. 
      The broken line on the boundary means to take  
   periodic boundary conditions, and the solid line on the 
   boundary means to take hard-wall boundary conditions.
   }
\vspfigB
\label{pw}\end{figure}

   The first case is the quasi-one-dimensional system 
with periodic boundary conditions in the $x$-direction 
and with hard-wall boundary conditions in the $y$-direction 
(the boundary case [\textit{A2}]). 
   A schematic illustration of this system is given  
in Fig. \ref{pw} in which periodic boundary conditions 
and hard-wall boundary conditions are 
represented as bold solid lines and broken lines, respectively. 

\begin{figure}[!htb]
\vspfigA
\caption{
   Stepwise structure of the Lyapunov spectrum  normalized 
   by the maximum Lyapunov exponent 
   for the quasi-one-dimensional system 
   with periodic boundary conditions in the $x$-direction 
   and hard-wall boundary conditions in the $y$-direction. 
   Inset: Full scale of the normalized Lyapunov spectrum.
   }
\vspfigB
\label{pwLya}\end{figure}  

   Figure \ref{pwLya} is the Lyapunov spectrum normalized 
by the maximum Lyapunov exponent $\lambda_{1}\approx 1.30$ 
for this system. 
   In this figure we showed a small positive region of the 
Lyapunov spectrum including its stepwise structures, while 
the full scale of the positive branch of the Lyapunov spectrum 
is shown in the inset. 
   In this system the $y$-component of the total momentum 
is not conserved because of the hard-wall boundary conditions 
in the $y$-direction, so there are only 4 zero-Lyapunov exponents 
in this system. 
   This figure shows clearly that the steps of the 
Lyapunov spectrum consist of four-point steps only, and 
there is no two-point step in the Lyapunov spectrum 
which appears in the model discussed in the previous section. 
   Besides, we cannot recognize a wavelike structure 
in the graph of Lyapunov vector components 
$\langle\delta q_{yj}^{(n)} \rangle$ as a function of 
the position $\langle q_{xj}\rangle$ (the graph [\textit{B1}]) 
in this model. 
   A comparison of this fact with results in the previous model 
suggests that the two-point step of the Lyapunov spectrum 
in the previous section should be strongly connected to 
the conservation of the $y$-component of the total momentum.

\begin{figure}[!htb]
\vspfigA
\caption{
      Local time-averaged quantities 
   $\langle\delta q_{yj}^{(n)}/p_{yj}\rangle_{t}$ 
   as functions of the normalized 
   position $\langle q_{xj} \rangle_{t}/L_{x}$ 
   and the collision number $n_{t}$
   corresponding to the Lyapunov exponents 
   $\lambda_{148}$ and $\lambda_{144}$ in the same collision 
   number interval $[390600,396000]$.
      The system is the quasi-one-dimensional system 
   with periodic boundary conditions in the $x$-direction 
   and hard-wall boundary conditions in the $y$-direction, 
   and the corresponding Lyapunov exponents are indicated 
   by arrows in Fig. \ref{pwLya}.
      Contour plots on the bottoms of these three-dimensional plots 
   are given by red dotted lines, black solid lines and 
   blue broken lines corresponding to the values 
   $\langle\delta q_{yj}^{(n)}/p_{yj} \rangle_{t} = 0.08, 
   0$ and $-0.08$, respectively.
   }
\vspfigB
\label{pwDqyPy}\end{figure}  

\begin{figure*}[!htb]
\vspfigA
\caption{
      Contour plots of the time-averaged quantities  
   $\langle\delta q_{yj }^{(n)}/p_{yj}\rangle_{t}$ 
   as functions of the normalized position 
   $\langle q_{xj}\rangle_{t}/L_{x}$ 
   and the collision number $n_{t}$ 
   corresponding to the first four-point step 
   consisting of the Lyapunov exponents $\lambda_{n}$, 
   $n=148,147,146$ and $145$ 
   in the same collision number interval 
   $[385800,396000]$.
      The system is the quasi-one-dimensional system 
   with periodic boundary conditions in the $x$-direction 
   and hard-wall boundary conditions in the $y$-direction, 
   and the corresponding Lyapunov exponents are indicated by the 
   brace under circles in Fig. \ref{pwLya}.
      Here red dotted lines, black solid lines and 
   blue broken lines correspond to the values 
   $\langle\delta q_{yj}^{(n)}/p_{yj} \rangle_{t} = 0.08, 
   0$ and $-0.08$, respectively. 
   }
\vspfigB
\label{pwDqyPyCont}\end{figure*}  

   We consider a relation between the four-point steps in 
the model of this section and in the model of the previous section 
by investigating the graph of the quantity  
$\langle\delta q_{yj}^{(n)}/p_{yj} \rangle_{t}$ 
as a function of the normalized position 
$\langle q_{xj}\rangle_{t}/L_{x}$ and 
the collision number $n_{t}$ (the graph [\textit{B2}]). 
   Fig. \ref{pwDqyPy} is such graphs, and  
corresponds to the Lyapunov exponent $\lambda_{148}$ 
in the first four-point step ( Fig. \ref{pwDqyPy}(a) )
and the Lyapunov exponent $\lambda_{144}$ in the second 
four-point step ( Fig. \ref{pwDqyPy}(b) ), 
which are indicated by arrows in Fig. 
\ref{pwLya}, in the same collision number interval 
$[390600,396000]$.
   The wavelike structures of these graphs have a wavelength 
$1/i$ for the $i$-th four-point steps.
   The time-oscillating period corresponding to 
the Lyapunov exponent $\lambda_{148}$ is almost the same as 
the period $\mathcal{T}_{0}$ of the first four-point steps 
of the previous model, and is approximately twice as long as  
the time-oscillating period in the Lyapunov exponent $\lambda_{144}$ 
of this model. 
   These features are common with the four-point steps in the 
models of the previous section, suggesting that the four-point steps of the 
Lyapunov spectrum in this model correspond to 
the four-point steps in the model of the previous section. 
   We can also show that the quantities 
$\langle\delta q_{yj}^{(n)}/p_{yj}\rangle_{t}$ 
corresponding to the zero Lyapunov exponents 
$\lambda_{150}$ and $\lambda_{149}$ are 
constants as functions of $\langle q_{xj}\rangle_{t}$ 
and $n_{t}$ approximately.

   Now we proceed to investigate the graph of the quantities  
$\langle\delta q_{yj}^{(n)}/p_{yj} \rangle_{t}$ 
corresponding to the Lyapunov exponents in 
the same four-point steps of the Lyapunov spectrum. 
   Fig. \ref{pwDqyPyCont} is the contour plots of 
these quantities as functions of the normalized position 
$\langle q_{xj}\rangle_{t}/L_{x}$ 
and the collision number $n_{t}$,  
corresponding to the first four-point step 
consisting of the Lyapunov exponents $\lambda_{n}$, 
$n=148,147,146$ and $145$, 
in the same collision number interval $[385800,396000]$. 
   The corresponding Lyapunov exponents are indicated by the 
brace under circles in Fig. \ref{pwLya}.
   In this case the contour 
lines of $\langle\delta q_{yj}^{(n)}/p_{yj}\rangle_{t} = 0$ 
seem to be slanting, but if we pay attention of the mountain 
regions and the valley regions in these graphs 
then we can realize that the structure of these 
four graphs are similar to the contour plots 
of the four graphs in Fig. \ref{ppDqyPyCont} for the previous model. 
   Therefore the phase relations among 
the time-oscillating wavelike structures of the quantity 
$\langle\delta q_{yj }^{(n)}/p_{yj}\rangle_{t}$ can be 
summarized in the schematic illustration given in Fig.  
\ref{phaseCont1} like in the previous model. 
   This suggests an approximate expression for the 
Lyapunov vector components given by Eq. (\ref{LyapuOscil}), 
for Lyapunov vector components $\delta q_{yj}^{(n)}$ 
corresponding to the Lyapunov exponents of the 
four-point steps in this model.

\subsection{The case of hard-wall boundary conditions in the 
   $x$-direction and periodic boundary conditions in the 
   $y$-direction}
\label{HarWalWP}

\begin{figure}[!htb]
\vspfigA
\includegraphics[width=\widthfigA]{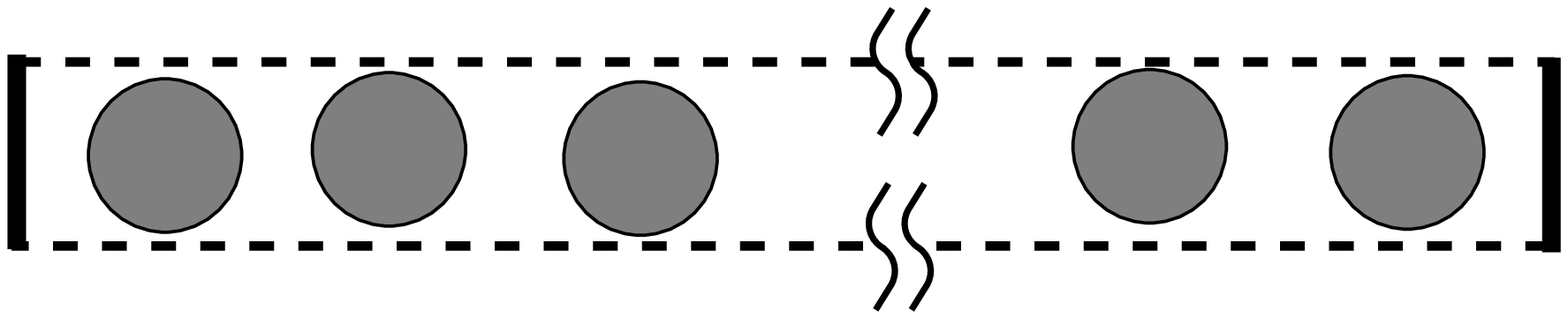}
\caption{
      A schematic illustration of a quasi-one-dimensional system 
   with hard-wall boundary conditions in the $x$-direction and 
   periodic boundary conditions in the $y$-direction. 
      The solid lines for the 
   boundary means to take hard-wall boundary conditions, 
   and broken lines for the boundary means to take 
   periodic boundary conditions.
   }
\vspfigB
\label{wp}\end{figure}

   As the next system we consider a quasi-one-dimensional system 
with hard-wall boundary conditions in the $x$-direction 
and periodic boundary conditions in the $y$-direction 
(the boundary case [\textit{A3}]). 
   A schematic illustration of such a system is given in Fig. 
\ref{wp} with solid lines for the hard-wall boundary 
conditions and broken lines for the periodic boundary conditions.
 
   In this system the $x$-component of the total momentum 
is not conserved, so the total number of the zero-Lyapunov 
exponents is 4. 
   Fig. \ref{wpLya} is a small positive region 
of the Lyapunov spectrum normalized by the maximum Lyapunov 
exponent $\lambda_{1}\approx 1.30$, 
including its stepwise structure, 
while its whole positive part of the Lyapunov spectrum is given 
in the inset.  
   The stepwise structure of the Lyapunov spectrum is 
clearly different from the model of the previous subsection, 
and consists of two-point steps interrupted by isolated 
single Lyapunov exponents. 
   (We call the isolated single Lyapunov exponents 
interrupting the two-point steps the "one-point steps" 
in this paper from now on, partly because they are  
connected to the two-point steps in the model of 
Sect. \ref{PerioPP} as will shown in this subsection.) 
   As discussed in Sect. \ref{PerioPP} the Lyapunov spectrum for 
the model with periodic boundary conditions in both  
directions has two-point steps and four-point steps,     
and it is remarkable that to adopt the hard-wall boundary 
conditions in 
the $x$-direction and to destroy the total momentum conservation in 
this direction halves the step-widths of both kinds of steps.

\begin{figure}[!htb]
\vspfigA
\caption{
      Stepwise structure of 
   the Lyapunov spectrum normalized by the maximum Lyapunov exponent 
   for the quasi-one-dimensional system 
   with hard-wall boundary conditions in the $x$-direction 
   and periodic boundary conditions in the $y$-direction. 
      Inset: Full scale of the normalized Lyapunov spectrum.
      The circle dots filled by black (gray) are 
   the Lyapunov exponents accompanying  
   wavelike structure (a constant behavior) 
   of the Lyapunov vector components  
   $\langle \delta q_{yj}^{(n)} \rangle$ partly shown in Fig. 
   \ref{wpDqy}.}
\vspfigB
\label{wpLya}\end{figure}  

   Fig. \ref{wpDqy} is the graphs of the time-averaged 
Lyapunov vector components  
$\langle \delta q_{yj}^{(n)} \rangle$ as functions of 
the normalized position $\langle q_{xj} \rangle/L_{x}$ 
(the graph [\textit{B1}]), corresponding to the one-point steps. 
   We can clearly recognize wavelike structures in these graphs, 
and in this sense the one-point steps in this model 
should be strongly related to the two-point steps in the 
model of Sect. \ref{PerioPP}. 
   The Lyapunov exponents accompanying wavelike structure 
of this kind of graphs are shown in Fig. \ref{wpLya}  
as black-filled circle dots, 
meaning that they are the one-point steps.  
   In Fig. \ref{wpLya} we also filled the circle dots 
by gray for 
one of the zero-Lyapunov exponents in which the graph of 
the Lyapunov vector components 
$\langle \delta q_{yj}^{(n)} \rangle$ as a function of 
the position 
$\langle q_{xj} \rangle$ is constant approximately.  
   However it is important to note that the wavelength 
of the wave corresponding to the $i$-th one-point step 
in this model is $2/i$, 
not $1/i$ like in the model of Sect. \ref{PerioPP}. 
   We fitted the graphs corresponding to the Lyapunov exponents 
$\lambda_{n}$, $n=149, 148, 145$ and $142$ by the functions 
$y=\alpha_{149}, \alpha_{148} \cos(\pi x + \beta_{148}), 
\alpha_{145} \cos(2\pi x + \beta_{145}), 
\alpha_{142} \cos(3\pi x + \beta_{142})$ respectively, with 
$\alpha_{n}$ and $\beta_{n}$ as fitting parameters. 
   Here the values of the fitting parameters are 
chosen as $\alpha_{149} = 0.11547$, 
$(\alpha_{148},\beta_{148}) = (0.16213,-0.0089778)$, 
$(\alpha_{145},\beta_{145}) = (-0.16202,-0.01007)$ and 
$(\alpha_{142},\beta_{142} = (-0.15945,0.0089148)$.
   The graphs are very nicely fitted by a constant or the 
sinusoidal functions, and lead to the form  

\begin{eqnarray}
   \delta q_{yj}^{(n(k))}  \approx 
   \alpha_{k}' \cos\left(\frac{\pi k}{L_{x}} q_{xj}
	 + \beta_{k}' \right) ,   
\label{LyapuWaveWP}\end{eqnarray}

\noindent $j=1,2,\cdots,N$, 
of the Lyapunov vector component $\delta q_{yj}^{(n(k))}$ 
corresponding to the Lyapunov exponents $\lambda_{n(k)}$ 
in the $k$-th one-point step  
with constants $\alpha_{k}'$ and $\beta_{k}'$. 

\begin{figure}[!htb]
\vspfigA
\caption{
      Time-averaged Lyapunov vector components  
   $\langle \delta q_{yj}^{(n)} \rangle$ corresponding to the 
   $n$-th Lyapunov exponent $\lambda_{n}$,  
   $n=149,148,145$ and $142$, 
   as functions of the time-averaged particle position 
   $\langle q_{xj} \rangle/L_{x}$ normalized by the 
   system length $L_{x}$.  
      The system is the quasi-one-dimensional system 
   with hard-wall boundary conditions in the $x$-direction 
   and periodic boundary conditions in the $y$-direction, 
   and the corresponding Lyapunov exponents are shown as  
   the black- and gray-filled circles in Fig. \ref{wpLya}. 
      The numerical data are fitted by a constant and 
   sinusoidal functions. 
   }
\vspfigB
\label{wpDqy}\end{figure}  

   Now we investigate the remaining steps, namely the two-point 
steps of the Lyapunov spectrum.
   Corresponding to these two-point steps of the Lyapunov spectrum, 
the graphs of the quantity 
$\langle\delta q_{yj}^{(n)}/p_{yj} \rangle_{t}$ 
as functions of the normalized position 
$\langle q_{xj}\rangle_{t}/L_{x}$ and the collision number $n_{t}$ 
(the graph [\textit{B2}]) show spatial wavelike structures 
oscillating in time. 
   It is shown in Fig. \ref{wpDqyPy} for those graphs 
corresponding to Lyapunov exponents (indicated by 
arrows in Fig. \ref{wpLya}) in different 
two-point step, in the same collision number interval 
$[222000,232200]$.
   The spatial wavelength of the waves corresponding to 
the $i$-th two-point step is $2/i$, which 
is twice as long as the wavelength of waves of the four-point steps 
in the models in Sect. \ref{PerioPP} and the previous subsection. 
   The period of time-oscillation of the wave corresponding 
to the $i$-th two-point step of the Lyapunov spectrum is approximately 
given by $\mathcal{T}_{0}'/i$ with a constant $\mathcal{T}_{0}'$. 
   This kind of graph corresponding to one of the zero-Lyapunov 
exponents, namely $\lambda_{150}$, is almost constant.

\begin{figure}[!htb]
\vspfigA
\caption{
      Local time-averaged quantities 
   $\langle\delta q_{yj}^{(n)}/p_{yj}\rangle_{t}$ 
   as functions of the normalized position 
   $\langle q_{xj}\rangle_{t}/L_{x}$ and the collision number 
   $n_{t}$  corresponding to the Lyapunov exponents $\lambda_{n}$, 
   $n=147,144$ and $141$ 
   in the same collision number interval $[222000,232200]$. 
      The system is the quasi-one-dimensional system 
   with hard-wall boundary conditions in the $x$-direction 
   and periodic boundary conditions in the $y$-direction, 
   and the corresponding Lyapunov exponents are indicated 
   by arrows in Fig. \ref{wpLya}. 
      Contour plots on the bottoms of these three-dimensional plots 
   are given by red dotted lines, black solid lines and 
   blue broken lines corresponding to the values 
   $\langle\delta q_{yj}^{(n)}/p_{yj} \rangle_{t} = 0.08, 
   0$ and $-0.08$, respectively.}
\vspfigB
\label{wpDqyPy}\end{figure}  

   It is important to note a relation between the 
time-oscillating period $\mathcal{T}_{0}$ of the 
preceding two models and the time-oscillating period 
$\mathcal{T}_{0}'$ of the model in this section. 
   Noting that in Figs. \ref{ppDqyPy}(a), \ref{pwDqyPy}(a) 
and \ref{wpDqyPy}(a) we plotted about one period of the 
time-oscillation of the wavelike structures of the quantities  
$\langle\delta q_{yj }^{(n)}/p_{yj}\rangle_{t}$ in the collision 
time intervals [543000,548100], [390600,396000] and 
[222000,232200], respectively, 
we can get an approximate relation $\mathcal{T}_{0}' 
\approx 2\mathcal{T}_{0}$.  

\begin{figure}[!htb]
\vspfigA
\caption{
   Contour plots of the local time-averaged quantities  
   $\langle\delta q_{yj }^{(n)}/p_{yj}\rangle_{t}$ 
   as functions of the normalized 
   position $\langle q_{xj}\rangle_{t}/L_{x}$ 
   and the collision number $n_{t}$ 
   corresponding to the first two-point step 
   consisting of the Lyapunov exponents $\lambda_{n}$, 
   $n=147$ and $146$ in the same collision number interval 
   $[212400,232200]$. 
   The system is the quasi-one-dimensional system 
   with hard-wall boundary conditions in the $x$-direction 
   and periodic boundary conditions in the $y$-direction,  
   and the corresponding Lyapunov exponents are indicated by the 
   brace under circles in Fig. \ref{wpLya}.
      Here red dotted lines, black solid lines and 
   blue broken lines correspond to the values 
   $\langle\delta q_{yj}^{(n)}/p_{yj} \rangle_{t} = 0.08, 
   0$ and $-0.08$, respectively. 
   }
\vspfigB
\label{wpDqyPyCont}\end{figure}  

\begin{figure}[!htb]
\vspfigA
\includegraphics[width=\widthfigC]{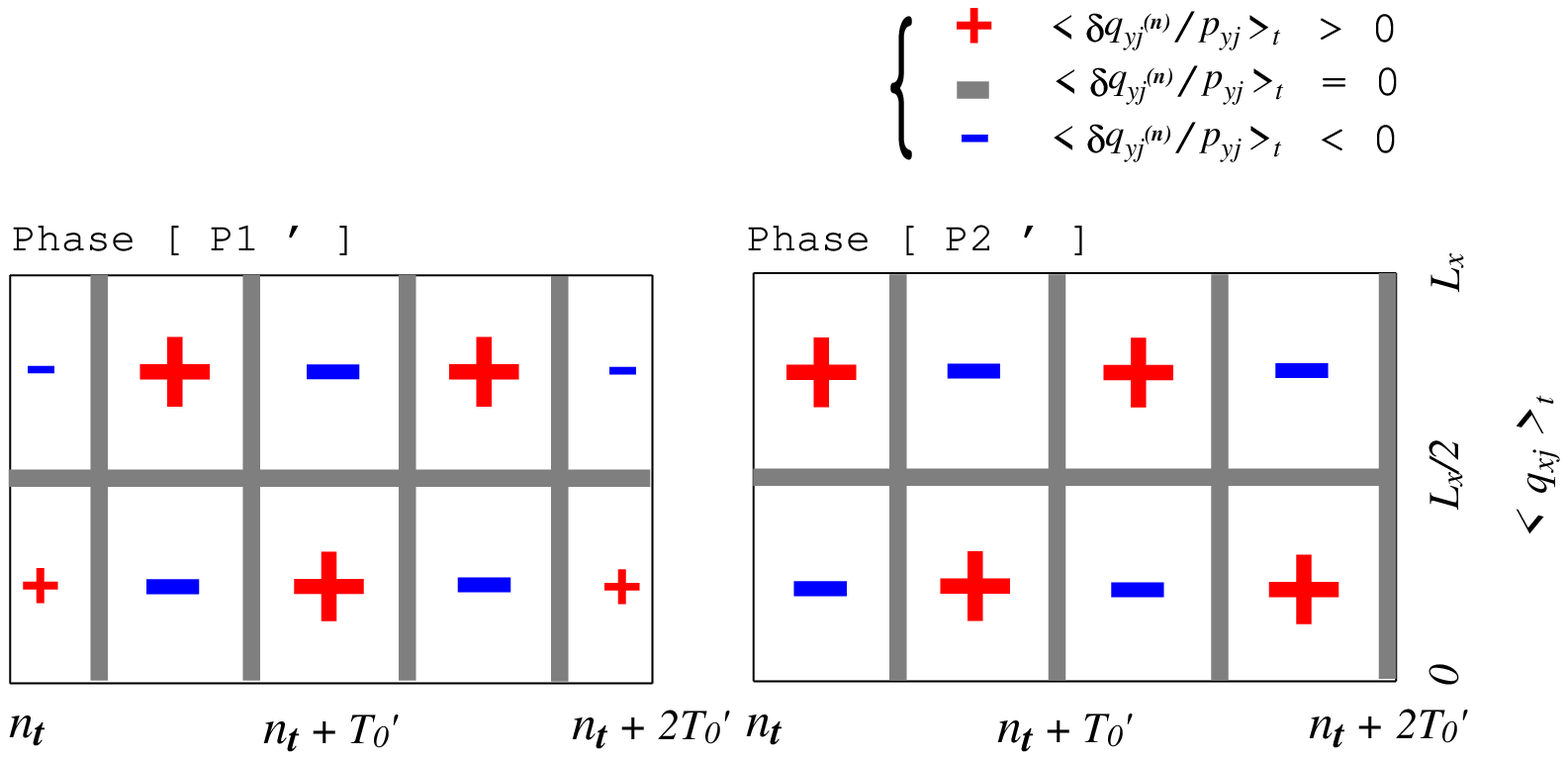}
\caption{
      Schematic illustration of the phase relations among 
   the time-oscillating wavelike structures of the quantities 
   $\langle\delta q_{yj}^{(n)}/p_{yj}\rangle_{t}$ 
   as functions of the position 
   $\langle q_{xj}\rangle_{t}$ 
   and the collision number $n_{t}$,  
   corresponding 
   to the first two-point step in the same collision number interval. 
      Thick gray lines means node lines, the region indicated by 
   a red plus sign ($+$) is the region where the quantity 
   $\langle\delta q_{yj }^{(n)}/p_{yj}\rangle_{t}$ is positive, 
   and the region indicated by 
   a blue minus sign ($-$) is the region where the quantity 
   $\langle\delta q_{yj }^{(n)}/p_{yj}\rangle_{t}$ is negative.
      The phase of [\textit{P1'}] is shifted in time from the phase 
   of [\textit{P2'}] by $\pi/2$. 
   }
\vspfigB
\label{phaseCont2}\end{figure}  

   The next problem is to investigate 
the graphs of the quantities 
$\langle\delta q_{yj}^{(n)}/p_{yj} \rangle_{t}$ 
as a function of the position 
$\langle q_{xj}\rangle_{t}$ and 
the collision number $n_{t}$ 
in the same two-point step of the Lyapunov spectrum. 
   Fig. \ref{wpDqyPyCont} is the contour plots of 
such graphs for the first two-point step
consisting of the Lyapunov exponents $\lambda_{n}$, 
$n=147$ and $146$, which is indicated by a  
brace under circles in Fig. \ref{wpLya}, in the same 
collision number interval $[212400,232200]$.
   It should be noted that the positions of 
the nodes of two spatial waves belonging to 
the same two-point step almost coincide with each other. 
   However the phases of the time-oscillations of the 
amplitudes of the waves are shifted by about $\pi/2$ 
with each other. 
   The phase relations of the graphs \ref{wpDqyPyCont}(a) 
and (b) is visualized in the schematic illustration given in Fig. 
\ref{phaseCont2} of  
the time-oscillating wavelike structures of the quantities  
$\langle\delta q_{yj }^{(n)}/p_{yj}\rangle_{t}$ corresponding 
to the first two-point step in the same collision number interval. 
   Here the phase [\textit{P1}'] and [\textit{P2}'] correspond 
to Figs. \ref{wpDqyPyCont}(a) and (b), respectively. 
   A similar investigation of the Lyapunov vectors shows that 
the time-oscillating wavelike structures for the second 
two-point steps are like the phases [\textit{P1}] and 
[\textit{P2}] of Fig. \ref{phaseCont1} except  
that the period $\mathcal{T}_{0}$ in 
Fig. \ref{phaseCont1} should be replaced 
with the oscillating period $\mathcal{T}_{0}'$ of this model. 
   These results suggest that the two-point steps in this model 
correspond to the four-point steps in the models of the previous 
subsection and Sect. \ref{PerioPP}, except for differences 
in the values of their wavelengths and time-oscillating periods. 
   After all we get a conjecture that the Lyapunov vector 
components $\delta q_{yj}^{(\tilde{n}(k))}$ and 
$\delta q_{yj}^{(\tilde{n}(k)-1)}$ corresponding to the 
Lyapunov exponents constructing the $k$-th two-point step 
are approximately expressed as 

\begin{eqnarray}
   && \hspace{-0.8cm} \left\{
      \delta q_{yj}^{(\tilde{n}(k))}, \; 
      \delta q_{yj}^{(\tilde{n}(k)-1)} 
   \right\} \approx 
   \nonumber \\
   \nonumber \\
   && \hspace{-0.2cm} \left\{
      \tilde{\alpha}_{k}' \; p_{yj}
	     \cos\left(\frac{\pi k}{L_{x}} q_{xj}
		    \!+\! \tilde{\beta}_{k}' \right) 
	     \cos\left( \frac{\pi k}{\mathcal{T}_{0}} n_{t}
            \!+\! \tilde{\gamma}_{k}' \right) , \right.   
    \nonumber \\
    &&  \left. \tilde{\alpha}_{k}' \; p_{yj}
	     \cos\left(\frac{\pi k}{L_{x}} q_{xj}
		    \!+\! \tilde{\beta}_{k}' \right) 
	     \sin\left( \frac{\pi k}{\mathcal{T}_{0}} n_{t}
            \!+\! \tilde{\gamma}_{k}' \right)
   \right\} ,  
\label{LyapuOscilWP}\end{eqnarray}

\noindent $j=1,2,\cdots,N$ 
with constants $\tilde{\alpha}_{k}'$, 
$\tilde{\beta}_{k}'$ and $\tilde{\gamma}_{k}'$,  
noting the time-oscillating period 
$\mathcal{T}_{0}'\approx 2\mathcal{T}_{0}$.



\subsection{The case of hard-wall boundary conditions 
   in both directions}
\label{HarWalWW}

\begin{figure}[!htb]
\vspfigA
\includegraphics[width=\widthfigA]{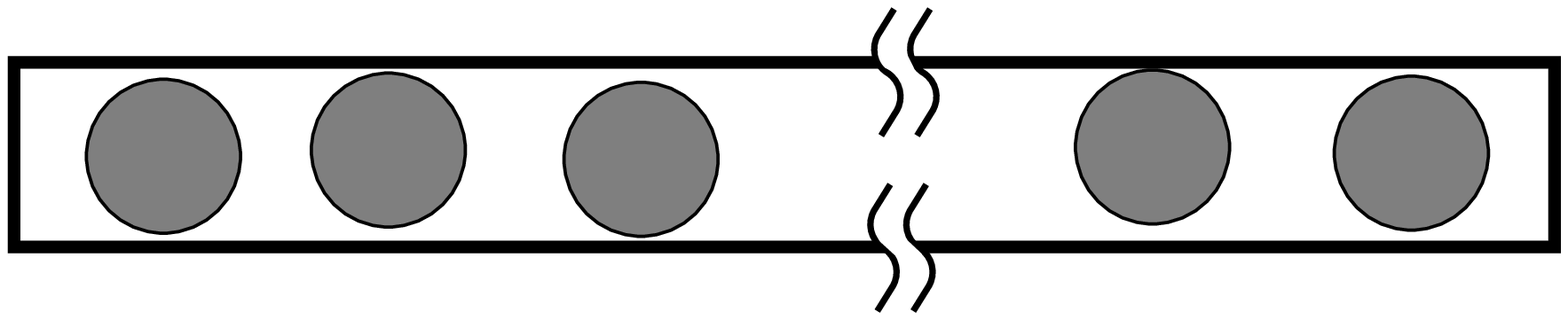}
\caption{
      A schematic illustration of a quasi-one-dimensional system 
   with hard-wall boundary conditions in both directions. 
      The solid line of the boundary means to take the 
   hard-wall boundary conditions.
   }
\vspfigB
\label{ww}\end{figure}

   The last model is the case of hard-wall boundary conditions 
in both directions  
(the boundary case [\textit{A4}]). 
   A schematic illustration of this system is given 
in Fig. \ref{ww} in which the solid line of the boundary means 
to take hard-wall boundary conditions. 

   A small positive region of the Lyapunov spectrum normalized by 
the maximum Lyapunov exponent $\lambda_{1}\approx 1.28$ 
is given in Fig. \ref{wwLya}. 
   The graph for the full scale of the positive branch of the 
normalized Lyapunov spectrum is also given in the inset  
of this figure. 
   In this system the total momentum is not conserved any more, 
and the total number of zero-Lyapunov exponents is 2.
   The stepwise structure of the Lyapunov spectrum consists of 
two-point steps only. 
   In this model a wavelike structure in the Lyapunov vector 
component $\langle\delta q_{yj}^{(n)} \rangle$ 
as a function of the position $\langle q_{xj}\rangle$ 
(the graph [\textit{B1}]) is not observed. 

\begin{figure}[!htb]
\vspfigA
\caption{
      Stepwise structure of 
   the Lyapunov spectrum normalized by the maximum Lyapunov exponent 
   for a quasi-one-dimensional system 
   with hard-wall boundary conditions in both directions. 
      Inset: Full scale of the normalized Lyapunov spectrum.
   }
\vspfigB
\label{wwLya}\end{figure}  

   Fig. \ref{wwDqyPy} is the graphs of the quantities 
$\langle\delta q_{yj}^{(n)}/p_{yj}\rangle_{t}$ 
as functions of the normalized position 
$\langle q_{xj}\rangle_{t}/L_{x}$ 
and the collision number $n_{t}$ 
(the graph [\textit{B2}]), corresponding to the Lyapunov 
exponents $\lambda_{n}$, 
$n=149,147$ and $145$, in the same collision number 
interval $[235200,246000]$. 
   The corresponding Lyapunov exponents are indicated 
by arrows in Fig. \ref{wwLya}. 
   This figure for the 
two-point steps of the Lyapunov spectrum shows a 
similar wavelike structure to 
the wavelike structure of the quantities 
$\langle\delta q_{yj}^{(n)}/p_{yj}\rangle_{t}$ 
in the model of the previous subsection, 
although one may think that fluctuations of these graphs 
in this model are much smaller than in the previous model. 
   This suggest that the two-point steps of the Lyapunov spectrum 
in Fig. \ref{wwLya} are similar to the two-point steps 
in the model of the previous subsection. 
   The wavelength of the spatial waves and the the periods  
of the time-oscillations corresponding to the $i$-th two-point 
step are $2/i$ and $\mathcal{T}_{0}'/i$, respectively, and 
the time-oscillating period of the first two-point 
step is given approximately 
by the same period $\mathcal{T}_{0}'$ $(\approx2\mathcal{T}_{0})$ 
as in the model of the previous subsection.   
   It may also be noted that this kind of graph corresponding to 
the zero-Lyapunov exponent $\lambda_{150}$ is almost constant.

\begin{figure}[!htb]
\vspfigA
\caption{
   Local time-averaged quantities 
   $\langle\delta q_{yj}^{(n)}/p_{yj}\rangle_{t}$ 
   as functions of the normalized 
   position $\langle q_{xj}\rangle_{t}/L_{x}$ and 
   the collision number $n_{t}$ corresponding to 
   the Lyapunov exponents $\lambda_{n}$, 
   $n=149,147$ and $145$ in the same collision number interval 
   $[235200,246000]$. 
      The system is the quasi-one-dimensional system 
   with hard-wall boundary conditions in both directions, 
   and the corresponding Lyapunov exponents are  
   indicated by arrows in Fig. \ref{wwLya}. 
      Contour plots on the bottoms of these three-dimensional plots 
   are given by red dotted lines, black solid lines and 
   blue broken lines corresponding to the values 
   $\langle\delta q_{yj}^{(n)}/p_{yj} \rangle_{t} = 0.08, 
   0$ and $-0.08$, respectively.
   }
\vspfigB
\label{wwDqyPy}\end{figure}  

   Fig. \ref{wwDqyPyCont} is the 
contour plots of the quantities  
$\langle\delta q_{yj}^{(n)}/p_{yj}\rangle_{t}$ 
as functions of the normalized position 
$\langle q_{xj}\rangle_{t}/L_{x}$  
and the collision number $n_{t}$ corresponding to 
the Lyapunov exponents $\lambda_{n}$, $n=149$ and $148$  
in the first two-point step of the Lyapunov spectrum 
in the same collision number interval 
$[223800,246000]$.
   The corresponding Lyapunov exponents are indicated by the 
brace under circles in Fig. \ref{wwLya}.
   Similarly to the previous model, 
the nodes of two spatial waves corresponding to 
the same two-point step almost coincide with each other, 
and the phase of the time-oscillation of the wave 
amplitudes is shifted by about $\pi/2$.  
   This also says that the phase relations of 
the graphs \ref{wwDqyPyCont}(a) 
and (b) is the types of the phases [\textit{P1}'] and 
[\textit{P2}'] in 
Fig. \ref{phaseCont2}. 
   In a similar way we can see that 
the time-oscillating wavelike structures for the second 
two-point steps of the Lyapunov spectrum for this model 
are like the phases [\textit{P1}] and 
[\textit{P2}] of Fig. \ref{phaseCont1} by 
replacing the period $\mathcal{T}_{0}$ of Fig. 
\ref{phaseCont1} with 
the time-oscillating period $\mathcal{T}_{0}'\approx 2\mathcal{T}_{0}$ 
of this model. 
   These suggest an approximate expression for the 
Lyapunov vector components given by Eq. (\ref{LyapuOscilWP}), 
for Lyapunov vector components $\delta q_{yj}^{(n)}$ 
corresponding to the Lyapunov exponents of 
the $k$-th two-point steps in this model.  
   The fact that the only difference between the model in this 
subsection and the model in the previous subsection is the 
boundary conditions in the $y$-direction suggests 
that the one-point steps of the model in the previous 
subsection come from the conservation of the $y$-component 
of the total momentum. 

\begin{figure}[!htb]
\vspfigA
\caption{
      Contour plots of the local time-averaged quantities 
   $\langle\delta q_{yj}^{(n)}/p_{yj}\rangle_{t}$ 
   as functions of the normalized 
   position $\langle q_{xj}\rangle_{t}/L_{x}$ 
   and the collision number $n_{t}$ 
   corresponding to the first two-point step 
   consisting of the Lyapunov exponents $\lambda_{n}$, 
   $n=149$ and $148$ in the same collision number interval 
   $[223800,246000]$.
      The system is the quasi-one-dimensional system 
   with hard-wall boundary conditions in both directions, 
   and the corresponding Lyapunov exponents are indicated by the 
   brace under circles in Fig. \ref{wwLya}.
      Here, red dotted lines, black solid lines and 
   blue broken lines correspond to the values 
   $\langle\delta q_{yj}^{(n)}/p_{yj} \rangle_{t} = 0.08, 
   0$ and $-0.08$, respectively. }
\vspfigB
\label{wwDqyPyCont}\end{figure}  



\newcommand{\widthtable}{1cm} 
\vspfigA
\begin{table*}[htbp]
\caption{
      The stepwise structures of the Lyapunov spectra 
   and the associated wavelike structures of the Lyapunov vectors in 
   the four boundary cases considered in this paper. 
      The boundary cases [\textit{A1}], [\textit{A2}], [\textit{A3}] 
   and [\textit{A4}] were considered 
   in the sections \ref{PerioPP}, \ref{HarWalPW}, 
   \ref{HarWalWP} and \ref{HarWalWW}, respectively.
      In this table, $\mathcal{S}$ is the number of points in the step 
   (or the number of the zero Lyapunov exponents in the line 
   specified by the label "$\lambda_{n} = 0$"), 
   $\mathcal{L}$ is the wavelength of spatial wavelike structure, 
   and $\mathcal{T}$ is the period of time-oscillation 
   of the wave,   
   $L_{x}$ is the length of the quasi-one-dimensional 
   rectangle, and $\mathcal{T}_{0}$ is constant.}
\begin{center}
\begin{tabular}{|c| |c|c|c| |c|c|c| |c|c|c| |c|c|c|} \hline
   & \multicolumn{3}{c||}{Boundary case [\textit{A1}]}  
   & \multicolumn{3}{c||}{Boundary case [\textit{A2}]}
   & \multicolumn{3}{c||}{Boundary case [\textit{A3}]}  
   & \multicolumn{3}{c|}{Boundary case [\textit{A4}]}  \\ \hline
    mode & \makebox[\widthtable]{$\mathcal{S}$} 
	& \makebox[\widthtable]{$\mathcal{L}$} 
	& \makebox[\widthtable]{$\mathcal{T}$} 
    & \makebox[\widthtable]{$\mathcal{S}$} 
    & \makebox[\widthtable]{$\mathcal{L}$} 
	& \makebox[\widthtable]{$\mathcal{T}$} 
    & \makebox[\widthtable]{$\mathcal{S}$} 
	& \makebox[\widthtable]{$\mathcal{L}$} 
	& \makebox[\widthtable]{$\mathcal{T}$} 
    & \makebox[\widthtable]{$\mathcal{S}$} 
    & \makebox[\widthtable]{$\mathcal{L}$} 
	& \makebox[\widthtable]{$\mathcal{T}$} \\ \hline
   $\lambda_{n} = 0$            
   & 6 & - & - & 4 & - & - 
   & 4 & - & - & 2 & - & - \\
   $\delta q_{yj}^{(n)}$        
   & 2 & $L_{x}$/1 & - & - & - & - 
   & 1 & 2$L_{x}$/1 & -  & - & - & - \\
   $\delta q_{yj}^{(n)}/p_{yj}$ 
   & 4 & $L_{x}$/1 & $\mathcal{T}_{0}$/1 & 4 & $L_{x}$/1 
      & $\mathcal{T}_{0}$/1 
   & 2 & 2$L_{x}$/1 & $2\mathcal{T}_{0}$/1 & 2 & 2$L_{x}$/1 
      & $2\mathcal{T}_{0}$/1 \\ 
   $\delta q_{yj}^{(n)}$      
   & 2 & $L_{x}$/2 & - & - & - & - 
   & 1 & 2$L_{x}$/2 & - & - & - & - \\
   $\delta q_{yj}^{(n)}/p_{yj}$ 
   & 4 & $L_{x}$/2  & $\mathcal{T}_{0}$/2  
   & 4 & $L_{x}$/2  & $\mathcal{T}_{0}$/2 
   & 2 & 2$L_{x}$/2 & $2\mathcal{T}_{0}$/2 
   & 2 & 2$L_{x}$/2 & $2\mathcal{T}_{0}$/2 \\ 
   $\delta q_{yj}^{(n)}$      
   & $\vdots$ & $\vdots$ & $\vdots$ & $\vdots$ & $\vdots$ & $\vdots$
   & 1 & 2$L_{x}$/3 & -   
       & - & - & - \\ 
   $\delta q_{yj}^{(n)}/p_{yj}$ 
   & $\vdots$ & $\vdots$ & $\vdots$ & $\vdots$ & $\vdots$ & $\vdots$ 
   & 2 & 2$L_{x}$/3 & $2\mathcal{T}_{0}$/3 
   & 2 & 2$L_{x}$/3 & $2\mathcal{T}_{0}$/3 \\ 
   $\vdots$ 
   & $\vdots$ & $\vdots$ & $\vdots$ & $\vdots$ & $\vdots$ & $\vdots$
   & $\vdots$ & $\vdots$ & $\vdots$ & $\vdots$ & $\vdots$ & $\vdots$ 
   \\ \hline
\end{tabular}
\end{center}
\label{sumar}
\end{table*}
\vspfigB

\section{Conclusion and Remarks}
\label{Concl}

   In this paper we have discussed numerically the 
stepwise structure of the Lyapunov 
spectra and its corresponding wavelike structures for the Lyapunov 
vectors in many-hard-disk systems. 
   We concentrated on the quasi-one-dimensional system whose 
shape is a very narrow rectangle that does not allow exchange of 
disk positions. 
   In the quasi-one-dimensional system we can get 
a stepwise structure of the Lyapunov 
spectrum in a relatively small system, for example, 
even in a 10-particle system, whereas a fully two-dimensional system 
would require a much more particles. 
   In such a system we have considered the following two 
problems: 
[\textit{A}] How does the stepwise structure of the Lyapunov 
spectra depend 
on boundary conditions such as periodic boundary conditions 
and hard-wall boundary conditions? 
[\textit{B}] How can we categorize the stepwise structure of the 
Lyapunov spectra using the wavelike structure of the 
corresponding Lyapunov vectors? 
   To consider the problem [\textit{A}] also means 
to investigate the effects 
of the total momentum conservation law on the stepwise structure 
of the Lyapunov spectra. 
   In this paper we considered four types of the boundary 
conditions;  
[\textit{A1}] periodic boundary conditions in the $x$- and 
$y$-directions, 
[\textit{A2}] periodic boundary conditions in the $x$-direction and 
hard-wall boundary conditions in the $y$-direction, 
[\textit{A3}] hard-wall boundary conditions in the $x$-direction and 
periodic boundary conditions in the $y$-direction, and 
[\textit{A4}] hard-wall boundary conditions in the $x$- and 
$y$-directions, 
in which we took the $y$-direction as the narrow direction of 
the rectangular shape of the system.   
   With each boundary case we obtained different stepwise structures 
of the Lyapunov spectra. 
   In each boundary condition we also considered 
graphs of the following two quantities; 
[\textit{B1}] the $y$-component $\delta q_{yj}^{(n)}$ of the 
spatial coordinate part of the Lyapunov vector of the 
$j$-th particle corresponding to the Lyapunov exponent $\lambda_{n}$ 
as a function of the $x$-component $q_{xj}$ of the spatial 
component of the $j$-th particle, and 
[\textit{B2}] the quantity $\delta q_{yj}^{(n)}/p_{yj}$ 
with the $y$-component 
$p_{yj}$ of the momentum coordinate of the $j$-th particle as 
a function of the position $q_{xj}$ and the collision number $n_{t}$. 
   These quantities $\delta q_{yj}^{(n)}$ and 
$\delta q_{yj}^{(n)}/p_{yj}$ give constant values in some of 
zero-Lyapunov exponents, at least approximately. 
   We found that the steps of the Lyapunov spectra accompany 
a wavelike structure in the quantity $\delta q_{yj}^{(n)}$ 
or $\delta q_{yj}^{(n)}/p_{yj}$, depending on the kind of 
steps of the Lyapunov spectra. 
   A time-dependent oscillating behavior appears 
in the wavelike structure of the quantity 
$\delta q_{yj}^{(n)}/p_{yj}$, whereas the wavelike structure 
of the quantity $\delta q_{yj}^{(n)}$ is essentially stationary. 
   Fluctuations of these quantities 
$\delta q_{yj}^{(n)}$ and $\delta q_{yj}^{(n)}/p_{yj}$ disturb  
their clear oscillatory structures, so we took a time-average of these 
quantities 
(a local time-average for the quantity $\delta q_{yj}^{(n)}/p_{yj}$ 
because of its time-oscillating behavior, 
and a longer time-average for the quantity $\delta q_{yj}^{(n)}$ 
because it is much more stationary in time than the quantity 
$\delta q_{yj}^{(n)}/p_{yj}$) 
to get their dominant wavelike structures.    
   In Table. \ref{sumar} we summarize our results about 
the characteristics 
of the stepwise structures of the Lyapunov spectra 
and the wavelike structures of the Lyapunov vectors 
in the four boundary cases 
[\textit{A1}], [\textit{A2}], [\textit{A3}] and [\textit{A4}]. 
   In Lyapunov exponents in each step of the Lyapunov spectra 
the wavelike structures of the quantity $\delta q_{yj}^{(n)}$ 
or $\delta q_{yj}^{(n)}/p_{yj}$ are approximately orthogonal 
to each others in space (in the sense of Eq. (\ref{LyapuWave})), 
in space and time (in the sense of Eq. (\ref{LyapuOscil})) 
or in time (in the sense of Eq. (\ref{LyapuOscilWP})), 
and this fact suggests  
that the wavelike structure of these quantities are sufficient to 
categorize the stepwise structure of the Lyapunov spectra in 
the quasi-one-dimensional systems considered in this paper.

   Different from a purely two-dimensional model 
such as a square system in which each particle can collide 
with any other particle, 
in the quasi-one-dimensional model a separation 
between the stepwise region and the smoothly changing region 
of the Lyapunov spectrum is not clear. 
   In Ref. \cite{Tan02} this point was explained as being 
caused by the fact that particles interact only with the 
two nearest-neighbor particles, whereas in the purely 
two-dimensional low-density system particles can interact 
with more than two particles. 

   In this paper we considered 
the quasi-one-dimensional systems only. 
   However there should be many interesting other situations 
in which we can investigate structures of the Lyapunov spectrum 
and the Lyapunov vectors. 
   For example, we may investigate the effect of the 
rotational invariance of the system on such structures 
by considering a two-dimensional system with a circle shape. 
   One might also investigate the system 
in which the orbit is not deterministic any more,  
in order to know whether the deterministic orbit plays  
an important role in the stepwise structure of the 
Lyapunov spectrum or not. 
   It may also be important to investigate the
dependence of the stepwise structures of the Lyapunov spectra 
on the spatial dimension of the system, 
for example, to investigate any structure 
of the Lyapunov spectra for purely 
one- or three-dimensional many particle systems. 
   (Note that the quasi-one-dimensional systems considered in 
this paper are still two-dimensional systems in the sense of the 
phase space dimension.)
   These problems remain to be investigated in the future.


\section*{Acknowledgements}

   The authors wish to thank C. P. Dettmann and H. A. Posch for 
valuable comments to this work. 
   One of the author (T.T) acknowledges information 
on presentations of figures by T. Yanagisawa. 
   We are grateful for financial support for this work 
from the Australian Research Council.



\end{document}